\documentclass[journal=jacsat,manuscript=article]{achemso}

\usepackage[version=3]{mhchem} 
\usepackage{xcolor}
\usepackage{adjustbox}
\usepackage{graphics}

\title{High-throughput DFT-based discovery of next generation two-dimensional (2D) superconductors }%

\author{Daniel Wines}%
 \email{daniel.wines@nist.gov}
\affiliation{%
 Material Measurement Laboratory, National Institute of Standards and Technology,
Gaithersburg, Maryland 20899, USA 
}%

\author{Kamal Choudhary}
 \affiliation{%
 Material Measurement Laboratory, National Institute of Standards and Technology,
Gaithersburg, Maryland 20899, USA 
}%
 \alsoaffiliation{%
Theiss Research, La Jolla, CA, 92037, USA}

\author{Adam J. Biacchi}%

\affiliation{%
 Physical Measurement Laboratory, National Institute of Standards and Technology,
Gaithersburg, Maryland 20899, USA 
}%

\author{Kevin F. Garrity}%

\affiliation{%
 Material Measurement Laboratory, National Institute of Standards and Technology,
Gaithersburg, Maryland 20899, USA 
}%

\author{Francesca Tavazza}%

\affiliation{%
 Material Measurement Laboratory, National Institute of Standards and Technology,
Gaithersburg, Maryland 20899, USA 
}%
\begin{document}

\begin{abstract}

High-throughput density functional theory (DFT) calculations allow for a systematic search for conventional superconductors. With the recent interest in two-dimensional (2D) superconductors, we used a high-throughput workflow to screen over 1,000 2D materials in the JARVIS-DFT database and performed electron-phonon coupling calculations, using the McMillan-Allen-Dynes formula to calculate the superconducting transition temperature ($T_c$) for 165 of them. Of these 165 materials, we identify 34 dynamically stable structures with transition temperatures above 5 K, including materials such as W$_2$N$_3$, NbO$_2$, ZrBrO, TiClO, NaSn$_2$S$_4$, Mg$_2$B$_4$C$_2$ and the previously unreported Mg$_2$B$_4$N$_2$ ($T_c$ = 21.8 K). Finally, we performed experiments to determine the $T_c$ of selected layered superconductors (2H-NbSe$_2$, 2H-NbS$_2$, ZrSiS, FeSe) and discuss the measured results within the context of our DFT results. We aim that the outcome of this workflow can guide future computational and experimental studies of new and emerging 2D superconductors by providing a roadmap of high-throughput DFT data.

\end{abstract}


\maketitle

In the past decade, superconductivity in two-dimensional (2D) systems has attracted a great deal of attention due to its potential applications for nanoscale devices such as superconducting transistors, quantum interferometers and superconducting qubits \cite{app1,PhysRevB.79.134530,doi:10.1126/science.299.5609.1045,doi:10.1063/1.3521262,app2}. Since the significant experimental work of Zhang et al. in 2010, which demonstrated superconductivity up to 1.8 K in single-layer Pb on Si(111) \cite{app3}, the quest to synthesize and theoretically predict 2D superconducting materials has moved towards intrinsically 2D monolayers whose bulk counterparts are weakly bonded layered materials \cite{intro1,Brun_2016,Saito_2016}. On the experimental front, a superconducting transition has been predicted for a variety of alkali-decorated graphene layers, with transitions reported for K-intercalated few-layer graphene at 4.5 K \cite{k-graph}, as well as for Li- and Ca-intercalated graphene \cite{Tiwari_2017,doi:10.1073/pnas.1510435112,doi:10.1063/1.4817781}. In addition, superconductivity has been measured for a variety of few-layer and monolayer transition metal dichalcogenides (TMDs) with a critical temperature of 7.2 K for NbSe$_2$ \cite{PhysRevLett.28.299,nbse2-1,nbse2-2,nbse2-3}, 5.3 K for NbS$_2$ \cite{PhysRevLett.101.166407,Yan_2019}, 3 K for TiSe$_2$ \cite{tise2}, 2.2 K for TaS$_2$ \cite{tas2} and between 7 and 12 K for MoS$_2$ \cite{doi:10.1126/science.aab2277,doi:10.1126/science.1228006,mos2-1,mos2-2}. 

Using first-principles calculations, efforts have been made to screen 2D superconducting monolayers with higher $T_c$ including intrinsic 2D metals and doped 2D materials \cite{Shao_2014,doi:10.1063/1.4916100,PhysRevB.93.104511,PhysRevB.99.064513,mof}. Two dimensional boron allotropes (such as borophene) \cite{PhysRevB.95.024505,boro,doi:10.1126/science.aad1080,doi:10.1063/1.4963179,boro2} and carbon-boron based 2D materials such as B$_2$O \cite{b2o}, B$_2$C \cite{C2NR12018F}, LiBC \cite{PhysRevB.104.054504}, MgB$_2$ \cite{PhysRevB.96.094510}, Mo$_2$C \cite{C7TC00789B}, Mg$_2$B$_4$C$_2$ \cite{Mg2B4C2}, and W$_2$N$_3$ \cite{doi:10.1021/acs.nanolett.0c05125} have been predicted to have substantial critical temperatures. Specifically, Mg$_2$B$_4$C$_2$ and W$_2$N$_3$ have been predicted to have a $T_c$ of 47 K and 21 K respectively \cite{Mg2B4C2,doi:10.1021/acs.nanolett.0c05125}. Computational predictions of superconducting transition temperatures can be most useful as a precursor to direct more expensive and time consuming experimental synthesis and characterization. Previously, there have been several efforts to systematically discover superconducting materials which fall into certain materials classes, such as transition metals \cite{roberts1976survey}, A15 , B1 \cite{kihlstrom1985tunneling,stewart2015superconductivity}, AB$_2$ compounds \cite{ivanovskii2003band,buzea2001review,nagamatsu2001superconductivity}, cuprates \cite{plakida2010high}, iron-based compounds \cite{hosono2015iron}, hydrides \cite{shipley2020stability,liu2017potential,drozdov2015conventional} and many other classes of materials \cite{poole2013superconductivity,shipley2021high,yuan2019recent,nagamatsu2001superconductivity,rodriguez1990optical,subedi2013electron,duan2019ab,kolmogorov2010new,gao2010high,duan2014pressure}. 

To efficiently identify Bardeen–Cooper–Schrieffer (BCS) conventional superconductors \cite{cooper2010bcs, giustino2017electron} with high critical temperatures using computational methods, there are two main requirements: 1) a robust computational workflow, and 2) a curated database of materials with prior knowledge such as band structure, magnetic properties and electronic density of states. Using density functional theory perturbation theory (DFT-PT), the electron-phonon coupling (EPC) can be calculated and used to predict $T_C$ with reasonable accuracy for a variety of  materials\cite{giustino2017electron, PhysRevB.101.134511}. In this work, we designed a systematic, data-driven workflow to expedite the discovery of new and potentially high-$T_C$ 2D superconductors. We do so by combining several computational methods at various levels of cost and accuracy. We start with a BCS-inspired pre-screening for metallic, nonmagnetic materials with high electron density of states (DOS) at the Fermi-level ($N(0)$), using the existing Joint Automated Repository for Various Integrated Simulations (JARVIS) DFT database for 2D materials \cite{choudhary2020joint}. Additionally, we searched existing literature for 2D superconductors not already in the JARVIS-DFT database and added them as appropriate. We then applied our DFT-PT workflow to compute $T_C$ using the EPC and the McMillan-Allen-Dynes formula \cite{allen1975transition}, with low convergence settings initially (k-points, q-points). For the most ideal candidates, we performed additional convergence to increase the accuracy of our predictions.

A key component in achieving an efficient search for 2D superconducting materials was to utilize the JARVIS (\url{https://jarvis.nist.gov/}) \cite{choudhary2020joint} infrastructure, which is a collection of databases and tools to automate materials design using density functional theory, classical force-fields, machine learning and experiments. JARVIS-DFT is a density functional theory based database of over 60,000 bulk materials and over 1,000 2D and 2D-like materials. Material properties such as formation energy, band gap \cite{choudhary2018computational}, exfoliation energies \cite{choudhary2017high}, solar-cell efficiency \cite{choudhary2019accelerated}, spin-orbit spillage \cite{choudhary2021high,choudhary2019high,choudhary2020computational}, elastic tensors \cite{choudhary2018elastic}, dielectric tensors, piezoelectric tensors, infrared and Raman spectra \cite{choudhary2020high}, electric field gradients \cite{choudhary2020density}, accurate magnetic properties \cite{https://doi.org/10.48550/arxiv.2209.10379}, and superconducting transition temperature of bulk materials \cite{https://doi.org/10.48550/arxiv.2205.00060}, all with strict and careful DFT-convergence criteria \cite{choudhary2019convergence}.

BCS-theory \cite{bardeen1957theory} states that the attractive electron-electron interaction mediated by phonons results in Cooper pairs, which are bound states that are formed by two electrons with opposite spins and momenta. BCS-theory gives the relation between the Debye temperature ($\theta_D$), electronic DOS at Fermi level $N(0)$, electron-phonon interaction ($V$) and the superconducting transition temperature ($T_C$):

\begin{equation} 
T_c=1.14\theta_D \exp\bigg(-\frac{1}{N(0)V}\bigg) \label{eq:bcs}
\end{equation} 
$\theta_D$ is defined as \cite{anderson1963simplified}:

\begin{equation} 
\theta_D = \frac{h}{k_B} \bigg[\frac{3n N_a \rho}{4\pi M}\bigg]^\frac{1}{3} v_m
\end{equation} 
where $h$ is Planck's constant, $k_B$ is the Boltzmann constant, $n$ is the number of atoms per formula unit, $N_A$ is Avogadro constant, $\rho$ is the crystal structure’s density, $M$ is the molar mass, and $v_m$ is the average sound velocity obtained from the elastic tensor \cite{anderson1963simplified}.

In our previous work on bulk materials \cite{https://doi.org/10.48550/arxiv.2205.00060}, we used the $\theta_D$ obtained from finite-difference calculations of elastic tensors\cite{choudhary2018elastic} (JARVIS-DFT contains elastic tensors for over 17,000 bulk materials). However, because of the lack of periodicity in one direction for 2D materials, the calculation of the elastic constants requires more careful consideration (performing several calculations of biaxial strain and then polynomial fitting, to allow for buckling in the out-of-plane direction). For this reason, elastic constants have not been computed for the over 1,000 2D materials in JARVIS-DFT. As a result, we cannot use $\theta_D$ as a screening metric. Instead, we must independently consider $N(0)$ in Eq. \ref{eq:bcs}, and screen based on the fact that a large $N(0)$ will yield a large value for $T_c$. This limitation to the screening is compensated by adding additional screening metrics such as band gap and magnetic moment to the criteria. Instead, our initial screening is based on $N(0)$, the DOS at the Fermi-level, which is available in JARVIS-DFT. We additionally screened materials based on whether or not they were nonmagnetic, and added candidate 2D superconducting materials to the JARVIS-DFT database from searching the literature.

The JARVIS-DFT database is primarily populated by calculations made using the Vienna Ab initio Simulation Package (VASP) \cite{kresse1996efficient,kresse1996efficiency} software and the OptB88vdW  \cite{klimevs2009chemical} functional. The k-points are carefully converged in JARVIS-DFT with respect to total energy \cite{choudhary2019convergence}. We used these converged k-points in subsequent Quantum Espresso \cite{giannozzi2009quantum,giannozzi2020quantum} electron-phonon calculations.

The properties of conventional BCS superconductors are directly related to the EPC. EPC calculations can be performed by methods such as the interpolated/Gaussian broadening method \cite{wierzbowska2005origins}, tetrahedron method \cite{kawamura2014improved} and Wannier-based electron-phonon methods \cite{ponce2016epw}. In the interpolated method, the integration over k-points involves replacing the delta function with a smeared function which has a finite broadening width, where the broadening width must be converged to obtain accurate results \cite{wierzbowska2005origins}. In the tetrahedron method, the k-points are analytically integrated in tetrahedral regions covering the Brillouin zone with the piece-wise linear interpolation of a matrix element \cite{kawamura2014improved}. Similar to the justification of our previous work \cite{https://doi.org/10.48550/arxiv.2205.00060}, we used the interpolated method for the EPC calculations due to the fact that our results obtained with the interpolation method are computationally less expensive and more stable (less variability with respect to k and q-points) than the tetrahedron method. Since the interpolated method was used, the value for broadening was carefully converged for each material.

We performed the EPC calculations using DFT-PT \cite{baroni1987green,gonze1995perturbation} with the Quantum Espresso code \cite{giannozzi2009quantum}, the Perdew-Burke-Ernzerhof functional revised for solids (PBEsol) \cite{perdew2008restoring} and Garrity-Bennett-Rabe-Vanderbilt (GBRV) \cite{garrity2014pseudopotentials} pseudopotentials. The starting structures were from the JARVIS-DFT 2D database (relaxed with OptB88vdW), and we performed a full re-relaxation with Quantum Espresso. In the Quantum Espresso re-relaxation, the unit cell and atomic positions were allowed to relax to minimize the force and external pressure on the system. The energy convergence value between two consecutive steps was chosen to be $10\times 10^{-8}$ eV and a maximum pressure of 0.5 kbar was allowed on the cell. After re-relaxing the structures with Quantum Espresso (using PBEsol and GBRV potentials), we obtained similar results to the original structure from JARVIS relaxed with VASP using OptB88vdW (i.e. there was about an average $\approx$ 1 - 2 $\%$ difference in in-plane lattice constant between the two methods). Similar to our previous work \cite{https://doi.org/10.48550/arxiv.2205.00060}, we used a 610 eV (45 Ry) plane-wave cutoff since we find that higher values have minimal effect on the calculated EPC parameters. The EPC parameter is derived from spectral function ${\alpha}^2 F(\omega)$, which is given by:

\begin{equation} 
{\alpha}^2 F(\omega)=\frac{1}{2{\pi}N({\epsilon_F})}\sum_{qj}\frac{\gamma_{qj}}{\omega_{qj}}\delta(\omega-\omega_{qj})w(q)
\end{equation} 
where $\omega_{qj}$ is the mode frequency, $N({\epsilon_F})$ is the DOS at the Fermi level ${\epsilon_F}$, $\delta$ is the Dirac-delta function, $w(q)$ is the weight of the $q$ point,  $\gamma_{qj}$ is the linewidth of a phonon mode $j$ at wave vector $q$ and can be written as:

\begin{equation} 
\gamma_{qj}=2\pi \omega_{qj} \sum_{nm} \int \frac{d^3k}{\Omega_{BZ}}|g_{kn,k+qm}^j|^2 \delta (\epsilon_{kn}-\epsilon_F) \delta(\epsilon_{k+qm}-\epsilon_F)
\end{equation} 
In this equation, the integral is over the first Brillouin zone, $\epsilon_{kn}$ and $\epsilon_{k+qm}$ are the eigenvalues from DFT with wavevector $k$ and $k+q$ within the $n$th and $m$th bands respectively, and $g_{kn,k+qm}^j$ is the electron-phonon matrix element. The relation between $\gamma_{qj}$ and the mode EPC parameter $\lambda_{qj}$ is as follows:

\begin{equation} 
\lambda_{qj}=\frac {\gamma_{qj}}{\pi hN(\epsilon_F)\omega_{qj}^2}
\end{equation} 
The EPC parameter is now given by:

\begin{equation} 
\lambda=2\int \frac{\alpha^2F(\omega)}{\omega}d\omega=\sum_{qj}\lambda_{qj}w(q)
\end{equation} 
with $w(q)$ as the weight of a $q$ point.

The superconducting transition temperature, $T_C$ can be approximated by using the original McMillan-Allen-Dynes \cite{allen1975transition} equation:

\begin{equation}
T_c=\frac{\omega_{log}}{1.2}\exp[-\frac{1.04(1+\lambda)}{\lambda-\mu^*(1+0.62\lambda)}]\label{eq:mad}
\end{equation}
where
\begin{equation} 
\omega_{log}=\exp\bigg[\frac{\int d\omega \frac{\alpha^2F(\omega)}{\omega}\ln\omega}{\int d\omega \frac{\alpha^2F(\omega)}{\omega}}\bigg]
\end{equation} 
In Eq.~\ref{eq:mad}, the $\mu^*$ parameter is the effective Coulomb potential. Although this parameter can be calculated from first principles \cite{lee1995first}, $\mu^*$ generally varies over a relatively small range (such as 0.09 to 0.18). Similar to several other studies involving 2D materials \cite{b2o,C2NR12018F,Mg2B4C2}, we take $\mu^*=0.1$ when reporting our results from high-throughput screening.

\begin{figure}
\caption{A full schematic of the the high-throughput workflow used to identify high $T_c$ 2D superconductors. The number of materials at each stage of the workflow are given.}
\begin{center}
\includegraphics[width=12cm]{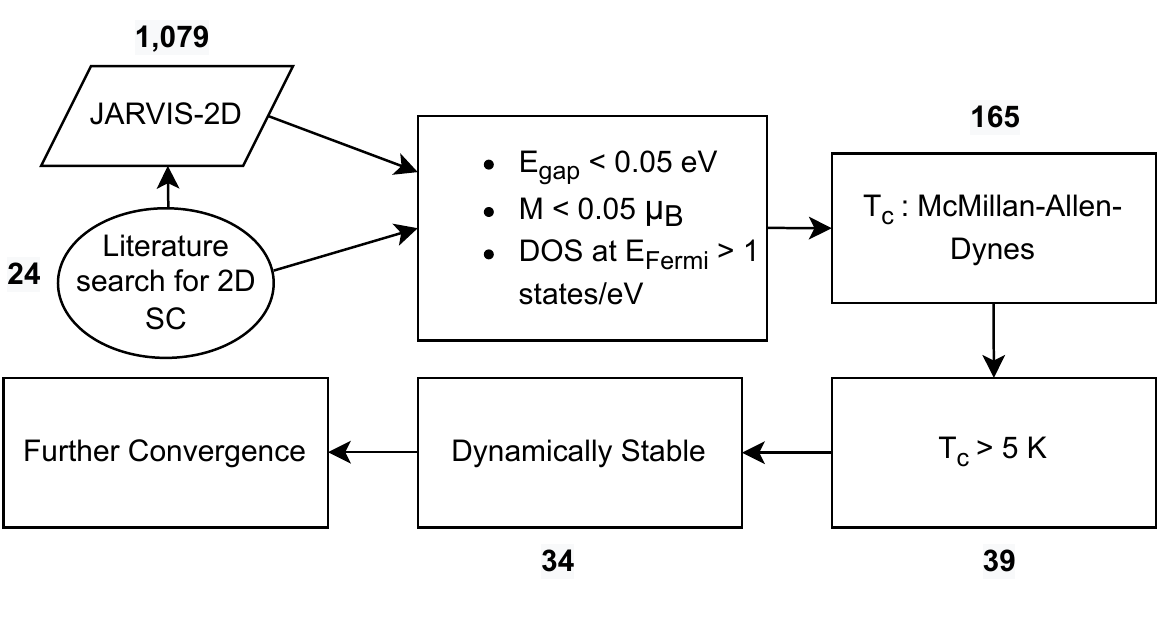}
\label{workflow}
\end{center}
\end{figure}

A full schematic of the high-throughput workflow we used to identify 2D superconductors with high $T_c$ is shown in Fig. \ref{workflow}. The starting point involves the screening of properties of the 1,079 monolayer materials to date in the JARVIS-DFT database. These 1,079 materials span a wide range of chemical and structural space, and their inclusion in the JARVIS-DFT database stems from studies involving high-throughput identification and characterization of 2D materials \cite{choudhary2017high}, investigation of elastic properties of 2D materials \cite{choudhary2018elastic}, the discovery of 2D solar cell materials \cite{choudhary2019accelerated}, discovery and characterization of 2D heterostructures \cite{CHOUDHARY20201}, and the high-throughput search for 2D topological materials \cite{choudhary2020computational}, thermoelectric materials \cite{Choudhary_2020} and anomalous quantum confinement materials \cite{PhysRevMaterials.5.054602}. Due to the fact that the elastic tensor (and therefore the Debye temperature) is more computationally expensive to compute for 2D materials than bulk materials, it is only available for a selected number of 2D materials in the JARVIS-DFT database. For this reason, we had to modify the BCS screening workflow that was used in our previous work \cite{https://doi.org/10.48550/arxiv.2205.00060}. Instead of screening the density of states (DOS) at the Fermi level and the Debye temperature (as we did for bulk materials), we screened materials based on the density of states (DOS) at the Fermi level, the total magnetic moment, and the electronic band gap (calculated with the OptB88vdW functional). This screening process is based on the notion that a potential 2D superconductor will have a high density of states at the Fermi level (and therefore be metallic), which is inspired by the BCS equation for $T_c$ (see Eq. \ref{eq:bcs}). In addition, for simplicity, we selected structures with zero magnetic moment per unit cell to avoid the magnetic moment interfering with the EPC. The EPC calculations carried out based on this screening are non-spin polarized. Based on this, we established a quantitative criteria for 2D superconductor screening depicted in Fig. \ref{workflow}.

In addition to selecting 2D materials in JARVIS that meet the criteria of E$_{\textrm{gap}}$ $<$ 0.05 eV, M $<$ 0.05 $\mu_{\textrm{B}}$, and DOS at E$_{\textrm{Fermi}}$ $>$ 1 states/eV/Nelect, we identified 24 additional 2D materials, based on an extensive literature search of 2D and 3D superconductors. These 24 2D materials were added to JARVIS-DFT and include: Mg$_2$B$_4$N$_2$, W$_2$N$_3$, MgB$_2$, Mo$_2$C, B$_2$N, Mg$_2$B$_4$C$_2$, $\chi$ and $\beta$ phase Borophene, ZrSiS, NbC, $\alpha$-Mg$_2$B$_4$N$_2$, B$_2$O, 2H-NbTe$_2$, 1T-NbTe$_2$, 2H-TaTe$_2$, 1T-TaTe$_2$, LiC$_6$, CaC$_6$, PdBi$_2$, W$_2$B$_3$, MgBH, ScC, Al$_4$CO and CaRu$_2$N$_2$. Although several of these 2D materials have been studied with ab-initio methods in previous works, to our knowledge, this is the first reported instance of 2D W$_2$B$_3$, Mg$_2$B$_4$N$_2$ (both phases), MgBH, NbC, ScC, and CaRu$_2$N$_2$.

In order to fully converge the EPC calculations with DFT-PT, a significant amount of computational resources are needed. These calculations generally require a dense k-point grid to sample electronic states and a dense q-point grid to sample phonons \cite{wierzbowska2005origins}. This becomes increasingly difficult for larger unit cells since the number of modes needed for calculation at each q-point increases as the number of atoms in the cell increases. In the supporting information, we display a number of convergence checks needed to determine the minimal set of parameters necessary to obtain a converged estimate of $T_c$. The quantities of interest include the number of k-points for the self consistent DFT calculation, the number of q-points for the DFT-PT calculation, the broadening value used for the linear interpolation, and the value for $\mu^*$ (in Eq. \ref{eq:mad}). In our previous work, we extensively investigated the effects of these parameters on the converged results \cite{https://doi.org/10.48550/arxiv.2205.00060}. 

We adopt a similar, less extensive procedure for selected 2D materials to confirm that our convergence criteria for bulk materials is valid in the 2D case. In Fig. S1 we show the convergence of $\lambda$, $\omega_{\textrm{log}}$ and $T_c$ with respect to different k-point and q-point grids with respect to broadening for the 2H-NbSe$_2$ compound, which is a well-known 2D superconductor experimentally with a high $T_c$ value. From Fig. S1) a) - c), we observe that as the q-point grid increases, the $T_c$ value has less of a dependence on the k-point grid. From analyzing all combinations of k-point and q-point grids, we calculate $T_c$ values that range from 4.5 K to 7.6 K. This is within reasonable agreement with respect to the experimental value of 7.2 K from literature \cite{PhysRevLett.28.299,nbse2-1,nbse2-2,nbse2-3} and we believe it is an acceptable margin of error for the high-throughput DFT-Allen-McMillan-Dynes method to obtain $T_c$. We also observe that a broadening parameter of 0.68 eV (0.05 Ry) is sufficient in converging the $T_c$ result within our desired tolerance, similar to our previous work \cite{https://doi.org/10.48550/arxiv.2205.00060}. This behavior is similar to the additional convergence tests we performed on a selected group of monolayers (Mg$_2$B$_4$N$_2$, W$_2$N$_3$, TiClO, NbO$_2$, 2H-NbS$_2$, ZrBrO, 2H-NbSe$_2$) shown in Fig. S2, which all display a converged $T_c$ value with respect to a broadening value of 0.68 eV (0.05 Ry). Additional q-point convergence (using the converged k-grid from JARVIS-DFT) for data is presented in Table S1 for 2D 2H-TaS$_2$, 2H-NbSe$_2$, W$_2$N$_3$, Mg$_2$B$_4$N$_2$ and TiClO. Based on these results, we determined that q-point grids as small as 2x2x1, combined with k-point grids similar in size to the typical grids used for self-consistent DFT total energy calculations (k-point
grid obtained from JARVIS-DFT total energy convergence \cite{choudhary2019convergence}) are sufficient to identify candidate 2D superconducting materials, along with a broadening of 0.68 eV (0.05 Ry). For selected promising candidate materials for which it was computationally feasible, we ran additional calculations with higher convergence parameters (see Table S2, which is an extension of Table \ref{tctable} to include the k-point and q-point grids used in the calculation).

\begin{figure}
\caption{Top and side view of the atomic structures of candidate 2D superconductors a) Mg$_2$B$_4$C$_2$, b) W$_2$N$_3$, c) 2H-NbS$_2$, d) 1T-NbS$_2$, e) TiClO, f) ZrBrO, g) NaSn$_2$S$_4$, h) $\chi$-Borophene, i) ZrSiS and j) FeSe. }
\begin{center}
\includegraphics[width=16cm]{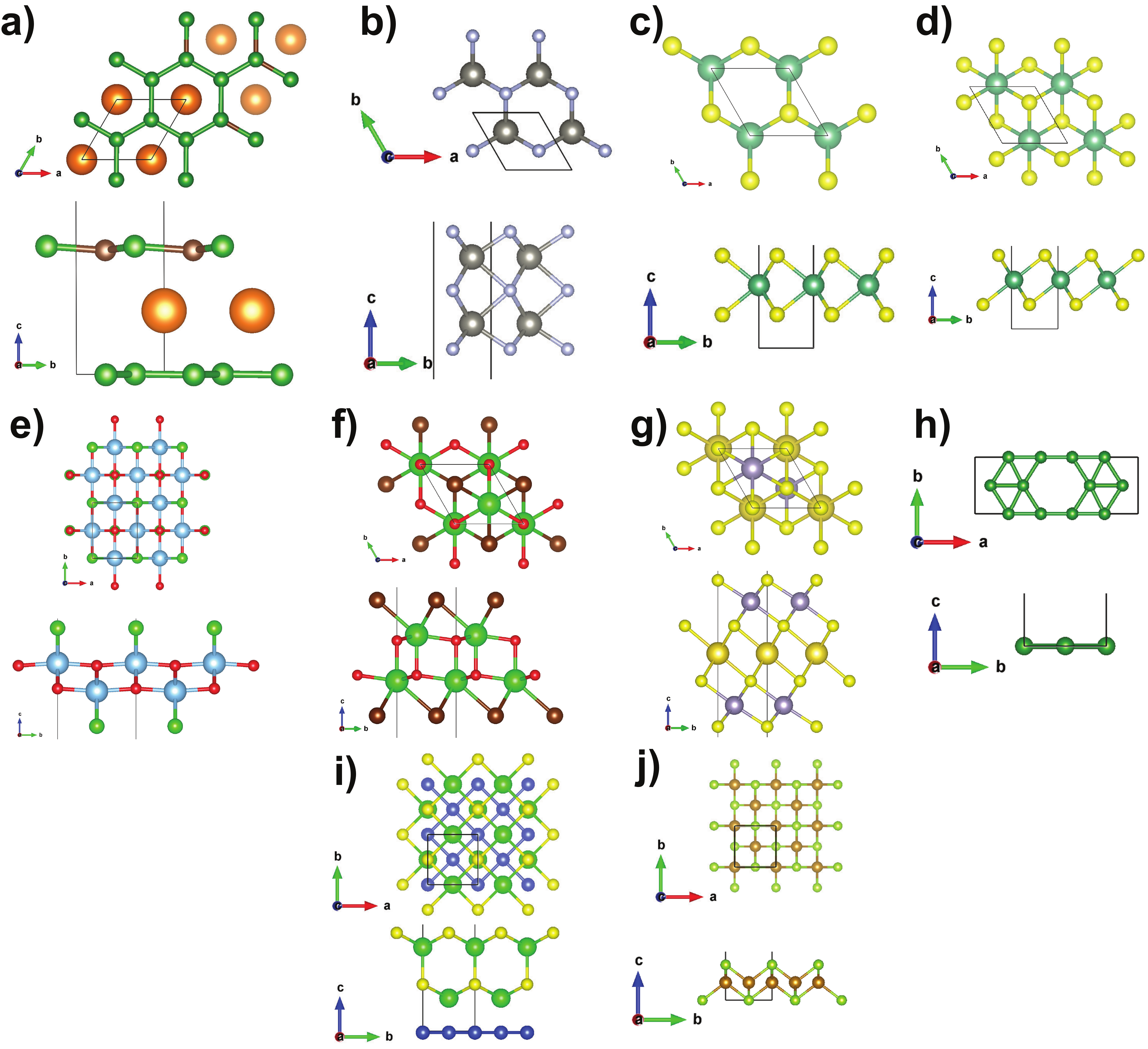}
\label{structurefig}
\end{center}
\end{figure}

In addition to the k-point, q-point and broadening convergence, we also addressed the impact of $\mu^*$ in our calculations \cite{giustino2017electron,wierzbowska2005origins,lee1995first,marques2005ab}. As previously mentioned, $\mu^*$ is the effective Coulomb potential in Eq. \ref{eq:mad}. Since it is computationally expensive to calculate this parameter from first principles, it is usually used as a fitting parameter for the calculation of $T_c$, and commonly taken as $\mu^*$ = 0.1. In Fig. S2 d), we plot the value of $T_c$ calculated with varying values of $\mu^*$, ranging from 0.03 to 0.22 for a selected group of representative superconductors (Mg$_2$B$_4$N$_2$, W$_2$N$_3$, TiClO, NbO$_2$, 2H-NbS$_2$, ZrBrO). We observe a remarkably consistent trend between all of these materials, with the $T_c$ values varying linearly with respect to the change in $\mu^*$. Across all six of these materials, we observe similar slopes (see S2 d)), with the slight exception of TiClO which has a smaller slope compared to the other materials due to the fact that it has a much larger value for $\lambda$ and a much smaller value for $\omega_{\textrm{log}}$. It is important to note that these six materials retain their superconducting properties despite a large range of $\mu^*$ values. The loss of superconducting properties at large values of $\mu^*$ has been reported for a large variety of bulk materials \cite{https://doi.org/10.48550/arxiv.2205.00060}. For the sake of being concise and consistent with previous works \cite{doi:10.1021/acs.nanolett.0c05125}, we report $T_c$ values calculated with $\mu^*$ = 0.1.

\begin{table}[]
\caption{\label{tctable} Tabulated results for the 2D superconductors with a $T_c$ above 5 K. Values for $\lambda$, $\omega_{\textrm{log}}$, $T_c$ and formation energy per atom are given. }
\begin{tabular}{l|l|l|l|l|l}
\hline
Structure & JARVIS ID            & $\lambda$ & $\omega_{\textrm{log}}$ & $T_c$ & E$_{\textrm{form}}$ \\
& & &  (K) & (K) & (eV/atom) \\
\hline
\hline
Mg$_2$B$_4$N$_2$            & JVASP-153112                     & 1.7                                 & 172.4                                 & 21.8                                                    & -0.24                              \\
W$_2$N$_3$               & JVASP-153122                  & 1.4                                 & 171.7                                 & 18.7                                                    & 0.05                               \\
NbO$_2$               & JVASP-31356                      & 1.6                                 & 144.5                                 & 17.5                                                    & -2.25                              \\
PtN$_2$               & JVASP-6634                       & 0.8                                 & 228.1                                 & 11.9                                                    & 1.59                               \\
MgB$_2$               & JVASP-153113                  & 0.6                                 & 472.2                                 & 11.8                                                    & 0.44                               \\
MoN                & JVASP-13586                      & 1.1                                 & 133.7                                 & 11.2                                                    & 1.90                               \\
TiClO              & JVASP-75097                       & 2.7                                 & 56.0                                  & 9.6                                                     & -2.42                              \\
B$_2$N                & JVASP-79276                    & 0.7                                 & 281.2                                 & 9.6                                                     & 0.09                               \\
ZrBrO              & JVASP-28185                     & 0.9                                 & 173.3                                 & 9.5                                                     & -2.39                              \\
2H-NbS$_2$             & JVASP-646                       & 1.1                                 & 117.4                                 & 9.3                                                     & -1.03                              \\
NaSn$_2$S$_4$            & JVASP-6949                    & 0.9                                 & 143.4                                 & 9.2                                                     & -0.60                              \\
Mg$_2$B$_4$C$_2$           & JVASP-153110                  & 0.5                                 & 655.5                                 & 9.0                                                     & -0.28                              \\
Nb$_2$CuS$_4$            & JVASP-75063               & 1.5                                 & 73.1                                  & 8.6                                                     & -0.52                              \\
AuN$_2$               & JVASP-75054                    & 1.4                                 & 80.6                                  & 8.4                                                     & 0.76                               \\
Nb$_2$CoS$_4$            & JVASP-27853       & 0.9        & 143.2                                 & 8.3                                                     & -0.59                              \\
1T-NbS$_2$             & JVASP-5947                    & 1.4                                 & 79.1                                  & 8.3                                                     & -1.00                              \\
$\chi$-Borophene              & JVASP-153104                    & 0.5                                 & 482.1                                 & 8.2                                                     & 0.48                               \\
NbC                & JVASP-153115                     & 0.8                                 & 197.9                                 & 8.1                                                     & 0.34                               \\
ZrSiS              & JVASP-153121                    & 0.9                                 & 128.2                                 & 7.9                                                     & -0.77                              \\
CoAs$_2$              & JVASP-6637                       & 1.3                                 & 80.0                                  & 7.8                                                     & 0.25                               \\
$\alpha$Mg$_2$B$_4$N$_2$            & JVASP-153111                     & 1.0                                 & 104.2                                 & 7.4                                                     & 0.06                               \\
B$_2$O                & JVASP-153100                 & 4.8                                 & 33.3                                  & 7.0                                                     & -0.73                              \\
2H-TaS$_2$            & JVASP-6070                    & 1.0                                 & 91.3                                  & 6.5                                                     & -1.06                              \\
2H-NbSe$_2$           & JVASP-655                      & 0.9                                 & 115.8                                 & 6.4                                                     & -0.76                              \\
BaSn$_4$O$_8$            & JVASP-77697                  & 1.3                                 & 64.0                                  & 6.3                                                     & -1.51                              \\
1T-NbSe$_2$            & JVASP-5899                      & 3.3                                 & 33.6                                  & 6.3                                                     & -0.74                              \\
LaBi$_2$O$_4$            & JVASP-28176                      & 1.4                                 & 58.5                                  & 6.2                                                     & -1.92                              \\
BrCY               & JVASP-60515                    & 1.0                                 & 86.8                                  & 6.0                                                     & -1.18                              \\
2H-NbTe$_2$            & JVASP-153106                     & 1.1                                 & 71.8                                  & 5.8                                                     & -0.33                              \\
TiSe               & JVASP-6010                      & 1.4                                 & 51.8                                  & 5.4                                                     & -0.74                              \\
TiS$_2$               & JVASP-774                        & 0.7                                 & 185.4                                 & 5.4                                                     & -1.33                              \\
ZrS                & JVASP-786                        & 0.8                                 & 123.3                                 & 5.3                                                     & -1.39                              \\
AuSe$_2$              & JVASP-6601                        & 2.7                                 & 29.9                                  & 5.1                                                     & 0.29                               \\
VSe                & JVASP-77610                       & 0.8                                 & 114.7                                 & 5.1                                                     & -0.47                             
\end{tabular}
\end{table}

\begin{figure}
\caption{a) and b) The relation between electron-phonon coupling parameters for all materials and the EPC function of some of the potential candidate superconductors c) W$_2$N$_3$, d) Mg$_2$B$_4$C$_2$, and e) 2H-NbSe$_2$. }
\begin{center}
\includegraphics[width=16.5cm]{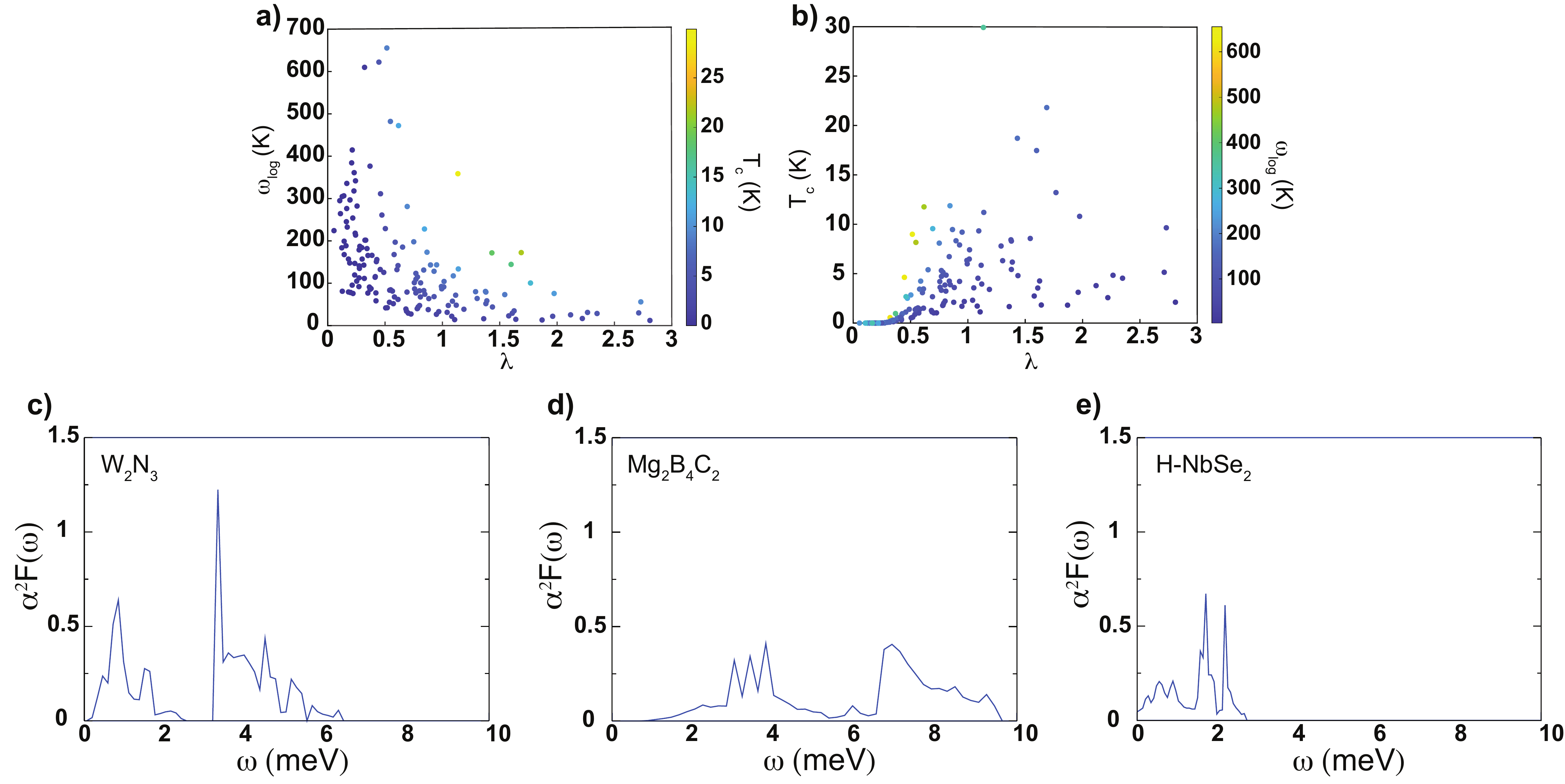}
\label{fullsc}
\end{center}
\end{figure}

We applied our high-throughput screening technique to the 1,079 monolayers in the JARVIS-DFT database. From this set of materials, and the 24 monolayers we added from literature, we found 165 materials that fit our criteria to perform the DFT-PT and McMillan-Allen-Dynes EPC calculation for. Similar to our previous work \cite{https://doi.org/10.48550/arxiv.2205.00060}, we ignore spin-orbit coupling, spin-polarization, Hubbard (+U) correction \cite{PhysRevB.57.1505}, and spin-fluctuation \cite{heid2010effect,gibson1989evidence} contributions due to the higher computational cost of such calculations. For a more accurate description of the underlying physics of these 2D superconductors, these effects should be considered in follow-up work. Of the 165 2D materials that we ran the EPC workflow for, we find that 34 have a $T_c$ greater than 5 K, and are dynamically stable (no imaginary phonon modes), which is an extremely promising result. It is also important to note that 11 of the 24 compounds that were added to JARVIS based on a literature search fit the criteria of a $T_c$ greater than 5 K and dynamic stability (included in the 34).

The tabulated results for these 34 materials are given in Table \ref{tctable}, where the chemical formula and EPC results are provided for each material ($\lambda$, $\omega_{\textrm{log}}$, $T_c$), together with their OptB88vdW-based formation energy per atom (E$_{\textrm{form}}$) from JARVIS-DFT. Out of these 34 materials, we ran the EPC calculation with the highest convergence parameters that were computationally tractable (k-point, q-point). The k-point and q-point grids used in the EPC calculations are given in Table S2. The JARVIS-IDs of each material (including the 24 newly added monolayers to JARVIS-DFT) are given in Table \ref{tctable} and Table S2. Each JARVIS-ID can be used to obtain more detailed information,
for example: https://www.ctcms.nist.gov/$\sim$knc6/static/JARVIS-DFT/JVASP-655.xml for JARVIS-ID: 655.          

We find that several candidate 2D superconductors that have a $T_c$ above 5 K are based on nitrides, borides and carbides (see Table \ref{tctable}). In addition, several oxide and niobium-based compounds and transition metal dichalcogenides such as NbS$_2$/Se$_2$ in both their 2H (two ``layers" per hexagonal unit cell) and 1T (one layer per trigonal unit cell) phases exhibit strong superconducting properties. Remarkably, 26 of these 34 materials have a formation energy less than or equal to 0.10 eV/atom and 23  of those 26 materials have negative formation energy, which signifies the likelihood that these materials can be synthesized and be stable in monolayer form. To further quantify this, we depict the phonon density of states for a selected number of structures with positive formation energy (see Fig. S3), where we see that although formation energy is positive, these structures are dynamically stable. In Fig. \ref{structurefig}, we depict the geometric structure of several of the promising candidate 2D superconductors for experimental synthesis. The JARVIS-ID of each material (which can be obtained from Table S2) can be used to obtain further properties and details. 

We observe the highest transition temperature for monolayer Mg$_2$B$_4$N$_2$, which has a $T_c$ of 21.8 K. To our knowledge, this material has previously been undiscovered in bulk and 2D form. The motivation to study Mg$_2$B$_4$N$_2$ stems from the recent prediction of a high superconducting temperature (47 K) \cite{Mg2B4C2} for Mg$_2$B$_4$C$_2$. In our work, we calculated the $T_c$ of 2D Mg$_2$B$_4$C$_2$ to be 9.0 K. The difference between the $T_c$ obtained in our work versus previous work can be attributed to the differences in calculation methods used (different codes, pseudopotentials, density functionals, method of perturbation theory/EPC calculation \cite{Mg2B4C2}). Nevertheless, the substitution of carbon with nitrogen results in over a 155 $\%$ increase in the superconducting transition temperature. Qualitatively this is an extremely promising result, and because it has negative formation energy, there is a strong motivation for experimentalists to pursue 2D Mg$_2$B$_4$N$_2$. Although we observe promising EPC properties for the Mg$_2$B$_4$N$_2$ monolayer, we observe that the B-N layers get slightly separated from the Mg-B after atomic relaxation. The atomic projected density of states indicates that contribution of $s$ and $p$ states of Mg and $p$ states of B is greatest around the Fermi level, while the atomic contribution of the B-N layers is away from the Fermi level. This implies that it is possible that there exists a hole-rich inner Mg-B slab in between two saturated B-N layers, where the EPC properties (high $T_c$) is coming mostly from the inner Mg-B layer. In addition, we found another previously unknown layered polymorph of Mg$_2$B$_4$N$_2$ (which we deem $\alpha$-Mg$_2$B$_4$N$_2$) with a high $T_c$ of 7.4 K. This hypothetical polymorph has a higher formation energy than the other Mg$_2$B$_4$N$_2$ structure (0.06 eV/atom vs. -0.24 eV/atom). Consistent with recent EPC calculations \cite{doi:10.1021/acs.nanolett.0c05125}, we find W$_2$N$_3$ to have a notably high $T_c$ of 18.7 K. Motivated by the investigation of Mg$_2$B$_4$N$_2$, we decided to substitute N with B in W$_2$N$_3$. We find a significant decrease in superconducting properties of the previously undiscovered W$_2$B$_3$ (JVASP-153120), with a calculated $T_c$ of 1.0 K. Our results on Mg$_2$B$_4$C$_2$/N$_2$ and W$_2$N$_3$/B$_3$ demonstrate how direct substitution can effectively tune the superconducting properties of a 2D material, which can be explored further in future work. We also investigated several 2D analogs of non-layered boron, carbon and nitrogen-based materials bulk materials (such as 2D MgB$_2$, B$_2$N, NbC, ScC, etc.). Although several of these monolayers have a high $T_c$ value (see Table \ref{tctable}), they have significantly positive formation energy, which makes their experimental realization less likely. In contrast, several 2D oxide-based materials (NbO$_2$, ZrBrO, TiClO, etc.) possess negative formation energy and strong superconducting properties.

\begin{figure}
\caption{Experimental zero field-cooled measurements of the DC magnetic susceptibility (using a magnetic field strength of 0.01 T) as a function of temperature, in order to determine $T_c$ for layered structures: a) 2H NbSe$_2$, b) 2H-NbS$_2$, c) FeSe, and d) ZrSiS.   }
\begin{center}
\includegraphics[width=15cm]{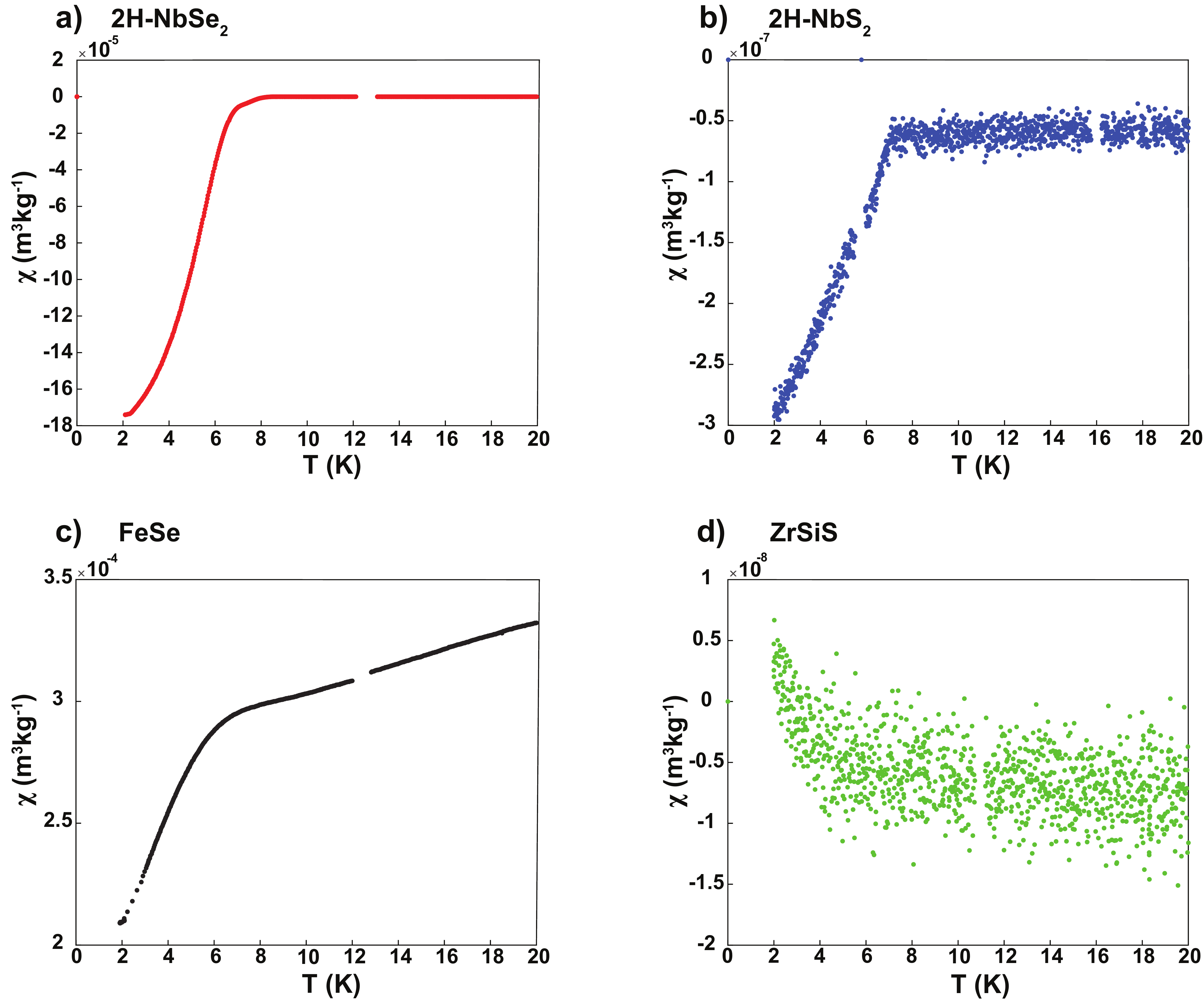}
\label{exp}
\end{center}
\end{figure}

In Fig. \ref{fullsc} a) and b), we show the relationship between EPC parameters for all of the considered 2D materials in this study. In Fig. \ref{fullsc} a) we observe that $\lambda$ and $\omega_{\textrm{log}}$ have an inverse relationship and in Fig. \ref{fullsc} b), we observe a positive relationship between $\lambda$ and $T_c$. Both of these behaviors are typical of BCS superconductors and were already observed in our work on BCS bulk superconductors \cite{https://doi.org/10.48550/arxiv.2205.00060}. As evident from the colormap of Fig. \ref{fullsc} a) and b), a balance of high $\lambda$ and $\omega_{\textrm{log}}$ is necessary for a material to have a high $T_c$. In Fig. \ref{fullsc} c) - e), we depict a few examples of the Eliashberg
spectral functions of some of the candidate materials (W$_2$N$_3$, Mg$_2$B$_4$C$_2$, and 2H-NbSe$_2$). The Eliashberg spectral function expresses the electron-phonon interaction in the form of a spectral density and the weighted area underneath the EPC function determines the $\lambda$ and $\omega_{\textrm{log}}$ parameters. These Eliashberg spectral functions and EPC parameters, along with several other material properties can be obtained from the JARVIS-DFT database.

In order to gain a deeper understanding of some of the materials we considered in this study, we performed zero field-cooled magnetometry experiments to determine the critical temperature. These measurements were done on powder samples, with the intention that they will be representative of the inherent 2D layered structure (although the discrepancy between single layer and bulk layered form is larger for certain materials). We conducted these experiments for layered 2H-NbSe$_2$, 2H-NbS$_2$, FeSe, and ZrSiS. Similar measurements have been conducted for 2H-NbSe$_2$, 2H-NbS$_2$, and FeSe, but to our knowledge, they have not been conducted for layered ZrSiS. Unfortunately, we could not conduct these zero field-cooled magnetometry experiments on some of the high $T_c$ materials such as W$_2$N$_3$, Mg$_2$B$_4$N$_2$, Mg$_2$B$_4$C$_2$ and several others due to the fact that they are not yet commercially available. Fig. \ref{exp} depicts the measured magnetic susceptibility (using a magnetic field strength of 0.01 T) as a function of temperature. We observe that out of these layered materials, 2H-NbSe$_2$ has a $T_c$ of 8.3 K, 2H-NbS$_2$ has a $T_c$ of 7.1 K, FeSe has a $T_c$ of 7.5 K, and ZrSiS does not have a superconducting transition due to the decreasing magnetic susceptibility with increasing temperature. The experimental results for layered 2H-NbSe$_2$ and 2H-NbS$_2$ are in excellent agreement (within $\approx$ 2 K) of the results we calculated for monolayer 2H-NbSe$_2$ and 2H-NbS$_2$ (see Table \ref{tctable}). Interestingly, layered ZrSiS (known to be a Dirac semimetal with topological properties \cite{zrsis-top,Aggarwal_2019}) does not possess superconducting properties, while our calculations for monolayer ZrSiS result in a high value of $T_c$ (7.9 K). The reason for this discrepancy stems from the major differences in the properties of the bulk layered structure vs. monolayer ZrSiS. In our previous work \cite{https://doi.org/10.48550/arxiv.2205.00060}, we calculated a $T_c$ of zero for bulk ZrSiS (JVASP-15288), which is consistent with the current measurements in Fig. \ref{exp} d). The presence of superconductivity in the monolayer and absence of superconductivity in the bulk structure is also consistent with previous experimental literature, where tip-induced superconductivity (with a $T_c$ of 7.5 K) was observed in single crystal ZrSiS \cite{Aggarwal_2019}. This can also be due in part to the fact that bulk ZrSiS has a smaller interlayer [vertical distance between consecutive layers in the bulk layered material, as opposed to the out of plane lattice constant $c$] distance (JVASP-15288, 2.2 \AA) compared to the interlayer distances of bulk NbS$_2$ (JVASP-30369, 3.4 \AA), NbSe$_2$ (JVASP-31795, 3.1 \AA), and FeSe (JVASP-45, 2.8 \AA). 

Finally, we decided to study layered FeSe due to the large amount of attention it has received in the past decade after superconductivity above 100 K was measured in single-layer FeSe films on a doped SrTiO$_3$ substrate \cite{fese}. Although we measure a high $T_c$ of 7.5 K for layered FeSe, we calculated a $T_c$ of 1.0 K for monolayer FeSe (JVASP-60244, the low $T_c$ is why it is absent from Table \ref{tctable}). We went on to calculate the EPC properties of bulk layered FeSe (JVASP-45) for comparison, and we find that it does not have a superconducting critical temperature (consistent with previous nonmagnetic DFT results for layered FeSe \cite{Zheng_2018}). The qualitative comparison between monolayer and bulk FeSe is significant in showing how the superconducting properties can be enhanced from the bulk to monolayer limit. The discrepancy between experiment and theory can be due in part to spin polarization not being taken into account. This is emphasized by previous computational studies where a substantial $T_c$ value was calculated for the antiferromagnetic orientation of FeSe \cite{Coh_2015}. These magnetic interactions of certain layered superconductors can play an important role in the calculation of accurate EPC parameters, but such calculations are beyond the scope of this high-throughput work and may be considered in future studies.

In summary, we have used a high-throughput DFT approach to study conventional 2D BCS superconductors, finding 34 candidate materials with a predicted $T_c$ above 5 K. Due to our high-throughput approach, we employ several approximations and assumptions, but we perform significant
benchmarking tests and convergence checks on particular materials to verify our results and methodology. We went one step further and performed experiments on selected layered superconductors to obtain the measured critical temperature. Our tools and datasets are publicly available (as a part of the JARVIS-DFT database) to enhance the transparency and reproducibility of this ongoing work. Possible applications of these additions to the JARVIS database include the data being used for training of various machine learning and generative models that can aid in the discovery of new superconductors. The calculated data can also be used for screening purposes for researchers who wish to tune the properties of promising 2D superconductors with methods such as applying pressure, alloying or creating heterostructures. We believe the results of this study can guide future computational and experimental studies of new and emerging 2D superconductors by providing a roadmap of high-throughput DFT data.

\section{Data Availability Statement}
The data from the present work can be found at https://figshare.com/articles/dataset/JARVIS-SuperconDB/21370572.

 \section{Code Availability Statement}
Software packages mentioned in the article can be found at https://github.com/usnistgov/jarvis. 

 \section{Notes}
Certain commercial equipment or materials are identified in this paper to adequately specify the experimental procedures.  In no case does the identification imply recommendation or endorsement by NIST, nor does it imply that the materials or equipment identified are necessarily the best available for the purpose. The authors declare no competing interests. Please note that the use of commercial software (VASP) does not imply recommendation by the National Institute of Standards and Technology.

 \section{Supporting Information}

Experimental methods, additional convergence details, additional details of dataset, phonon density of states, list of newly added structures to JARVIS.

\section{Acknowledgments}
All authors thank the National Institute of Standards and Technology for funding, computational, and data-management resources. K.C. thanks the computational support from XSEDE (Extreme Science and Engineering Discovery Environment) computational resources under allocation number TG-DMR 190095. Contributions from K.C. were supported by the financial assistance award 70NANB19H117 from the U.S. Department of Commerce, National Institute of Standards and Technology. The authors would like to thank Dr. Sobhit Singh for helpful discussions during the preparation of this manuscript.

\bibliography{Main}

\providecommand{\noopsort}[1]{}\providecommand{\singleletter}[1]{#1}%
\providecommand{\latin}[1]{#1}
\makeatletter
\providecommand{\doi}
  {\begingroup\let\do\@makeother\dospecials
  \catcode`\{=1 \catcode`\}=2 \doi@aux}
\providecommand{\doi@aux}[1]{\endgroup\texttt{#1}}
\makeatother
\providecommand*\mcitethebibliography{\thebibliography}
\csname @ifundefined\endcsname{endmcitethebibliography}
  {\let\endmcitethebibliography\endthebibliography}{}
\begin{mcitethebibliography}{107}
\providecommand*\natexlab[1]{#1}
\providecommand*\mciteSetBstSublistMode[1]{}
\providecommand*\mciteSetBstMaxWidthForm[2]{}
\providecommand*\mciteBstWouldAddEndPuncttrue
  {\def\EndOfBibitem{\unskip.}}
\providecommand*\mciteBstWouldAddEndPunctfalse
  {\let\EndOfBibitem\relax}
\providecommand*\mciteSetBstMidEndSepPunct[3]{}
\providecommand*\mciteSetBstSublistLabelBeginEnd[3]{}
\providecommand*\EndOfBibitem{}
\mciteSetBstSublistMode{f}
\mciteSetBstMaxWidthForm{subitem}{(\alph{mcitesubitemcount})}
\mciteSetBstSublistLabelBeginEnd
  {\mcitemaxwidthsubitemform\space}
  {\relax}
  {\relax}

\bibitem[De~Franceschi \latin{et~al.}(2010)De~Franceschi, Kouwenhoven,
  Sch{\"o}nenberger, and Wernsdorfer]{app1}
De~Franceschi,~S.; Kouwenhoven,~L.; Sch{\"o}nenberger,~C.; Wernsdorfer,~W.
  Hybrid superconductor--quantum dot devices. \emph{Nature Nanotechnology}
  \textbf{2010}, \emph{5}, 703--711\relax
\mciteBstWouldAddEndPuncttrue
\mciteSetBstMidEndSepPunct{\mcitedefaultmidpunct}
{\mcitedefaultendpunct}{\mcitedefaultseppunct}\relax
\EndOfBibitem
\bibitem[Huefner \latin{et~al.}(2009)Huefner, May, Kicin, Ensslin, Ihn, Hilke,
  Suter, de~Rooij, and Staufer]{PhysRevB.79.134530}
Huefner,~M.; May,~C.; Kicin,~S.; Ensslin,~K.; Ihn,~T.; Hilke,~M.; Suter,~K.;
  de~Rooij,~N.~F.; Staufer,~U. Scanning gate microscopy measurements on a
  superconducting single-electron transistor. \emph{Phys. Rev. B}
  \textbf{2009}, \emph{79}, 134530\relax
\mciteBstWouldAddEndPuncttrue
\mciteSetBstMidEndSepPunct{\mcitedefaultmidpunct}
{\mcitedefaultendpunct}{\mcitedefaultseppunct}\relax
\EndOfBibitem
\bibitem[Delahaye \latin{et~al.}(2003)Delahaye, Hassel, Lindell,
  Sillanp{\"a}{\"a}, Paalanen, Sepp{\"a}, and
  Hakonen]{doi:10.1126/science.299.5609.1045}
Delahaye,~J.; Hassel,~J.; Lindell,~R.; Sillanp{\"a}{\"a},~M.; Paalanen,~M.;
  Sepp{\"a},~H.; Hakonen,~P. Low-Noise Current Amplifier Based on Mesoscopic
  Josephson Junction. \emph{Science} \textbf{2003}, \emph{299},
  1045--1048\relax
\mciteBstWouldAddEndPuncttrue
\mciteSetBstMidEndSepPunct{\mcitedefaultmidpunct}
{\mcitedefaultendpunct}{\mcitedefaultseppunct}\relax
\EndOfBibitem
\bibitem[Romans \latin{et~al.}(2010)Romans, Osley, Young, Warburton, and
  Li]{doi:10.1063/1.3521262}
Romans,~E.~J.; Osley,~E.~J.; Young,~L.; Warburton,~P.~A.; Li,~W.
  Three-dimensional nanoscale superconducting quantum interference device
  pickup loops. \emph{Applied Physics Letters} \textbf{2010}, \emph{97},
  222506\relax
\mciteBstWouldAddEndPuncttrue
\mciteSetBstMidEndSepPunct{\mcitedefaultmidpunct}
{\mcitedefaultendpunct}{\mcitedefaultseppunct}\relax
\EndOfBibitem
\bibitem[Liu and Hersam(2019)Liu, and Hersam]{app2}
Liu,~X.; Hersam,~M.~C. 2D materials for quantum information science.
  \emph{Nature Reviews Materials} \textbf{2019}, \emph{4}, 669--684\relax
\mciteBstWouldAddEndPuncttrue
\mciteSetBstMidEndSepPunct{\mcitedefaultmidpunct}
{\mcitedefaultendpunct}{\mcitedefaultseppunct}\relax
\EndOfBibitem
\bibitem[Zhang \latin{et~al.}(2010)Zhang, Cheng, Li, Sun, Wang, Zhu, He, Wang,
  Ma, Chen, Wang, Liu, Lin, Jia, and Xue]{app3}
Zhang,~T.; Cheng,~P.; Li,~W.-J.; Sun,~Y.-J.; Wang,~G.; Zhu,~X.-G.; He,~K.;
  Wang,~L.; Ma,~X.; Chen,~X.; Wang,~Y.; Liu,~Y.; Lin,~H.-Q.; Jia,~J.-F.;
  Xue,~Q.-K. Superconductivity in one-atomic-layer metal films grown on
  Si(111). \emph{Nature Physics} \textbf{2010}, \emph{6}, 104--108\relax
\mciteBstWouldAddEndPuncttrue
\mciteSetBstMidEndSepPunct{\mcitedefaultmidpunct}
{\mcitedefaultendpunct}{\mcitedefaultseppunct}\relax
\EndOfBibitem
\bibitem[Saito \latin{et~al.}(2016)Saito, Nojima, and Iwasa]{intro1}
Saito,~Y.; Nojima,~T.; Iwasa,~Y. Highly crystalline 2D superconductors.
  \emph{Nature Reviews Materials} \textbf{2016}, \emph{2}, 16094\relax
\mciteBstWouldAddEndPuncttrue
\mciteSetBstMidEndSepPunct{\mcitedefaultmidpunct}
{\mcitedefaultendpunct}{\mcitedefaultseppunct}\relax
\EndOfBibitem
\bibitem[Brun \latin{et~al.}(2016)Brun, Cren, and Roditchev]{Brun_2016}
Brun,~C.; Cren,~T.; Roditchev,~D. Review of 2D superconductivity: the ultimate
  case of epitaxial monolayers. \emph{Superconductor Science and Technology}
  \textbf{2016}, \emph{30}, 013003\relax
\mciteBstWouldAddEndPuncttrue
\mciteSetBstMidEndSepPunct{\mcitedefaultmidpunct}
{\mcitedefaultendpunct}{\mcitedefaultseppunct}\relax
\EndOfBibitem
\bibitem[Saito \latin{et~al.}(2016)Saito, Nojima, and Iwasa]{Saito_2016}
Saito,~Y.; Nojima,~T.; Iwasa,~Y. Gate-induced superconductivity in
  two-dimensional atomic crystals. \emph{Superconductor Science and Technology}
  \textbf{2016}, \emph{29}, 093001\relax
\mciteBstWouldAddEndPuncttrue
\mciteSetBstMidEndSepPunct{\mcitedefaultmidpunct}
{\mcitedefaultendpunct}{\mcitedefaultseppunct}\relax
\EndOfBibitem
\bibitem[Xue \latin{et~al.}(2012)Xue, Chen, Yang, Zhu, Wang, He, and
  Cao]{k-graph}
Xue,~M.; Chen,~G.; Yang,~H.; Zhu,~Y.; Wang,~D.; He,~J.; Cao,~T.
  Superconductivity in Potassium-Doped Few-Layer Graphene. \emph{Journal of the
  American Chemical Society} \textbf{2012}, \emph{134}, 6536--6539\relax
\mciteBstWouldAddEndPuncttrue
\mciteSetBstMidEndSepPunct{\mcitedefaultmidpunct}
{\mcitedefaultendpunct}{\mcitedefaultseppunct}\relax
\EndOfBibitem
\bibitem[Tiwari \latin{et~al.}(2017)Tiwari, Shin, Hwang, Jung, Park, and
  Lee]{Tiwari_2017}
Tiwari,~A.~P.; Shin,~S.; Hwang,~E.; Jung,~S.-G.; Park,~T.; Lee,~H.
  Superconductivity at 7.4{\hspace{0.167em}}K in few layer graphene by
  Li-intercalation. \emph{Journal of Physics: Condensed Matter} \textbf{2017},
  \emph{29}, 445701\relax
\mciteBstWouldAddEndPuncttrue
\mciteSetBstMidEndSepPunct{\mcitedefaultmidpunct}
{\mcitedefaultendpunct}{\mcitedefaultseppunct}\relax
\EndOfBibitem
\bibitem[Ludbrook \latin{et~al.}(2015)Ludbrook, Levy, Nigge, Zonno, Schneider,
  Dvorak, Veenstra, Zhdanovich, Wong, Dosanjh, Stra{\ss}er, St{\"o}hr, Forti,
  Ast, Starke, and Damascelli]{doi:10.1073/pnas.1510435112}
Ludbrook,~B.~M. \latin{et~al.}  Evidence for superconductivity in Li-decorated
  monolayer graphene. \emph{Proceedings of the National Academy of Sciences}
  \textbf{2015}, \emph{112}, 11795--11799\relax
\mciteBstWouldAddEndPuncttrue
\mciteSetBstMidEndSepPunct{\mcitedefaultmidpunct}
{\mcitedefaultendpunct}{\mcitedefaultseppunct}\relax
\EndOfBibitem
\bibitem[Li \latin{et~al.}(2013)Li, Feng, Zhang, Ou, Chen, He, Wang, Guo, Liu,
  Xue, and Ma]{doi:10.1063/1.4817781}
Li,~K.; Feng,~X.; Zhang,~W.; Ou,~Y.; Chen,~L.; He,~K.; Wang,~L.-L.; Guo,~L.;
  Liu,~G.; Xue,~Q.-K.; Ma,~X. Superconductivity in Ca-intercalated epitaxial
  graphene on silicon carbide. \emph{Applied Physics Letters} \textbf{2013},
  \emph{103}, 062601\relax
\mciteBstWouldAddEndPuncttrue
\mciteSetBstMidEndSepPunct{\mcitedefaultmidpunct}
{\mcitedefaultendpunct}{\mcitedefaultseppunct}\relax
\EndOfBibitem
\bibitem[Frindt(1972)]{PhysRevLett.28.299}
Frindt,~R.~F. Superconductivity in Ultrathin Nb${\mathrm{Se}}_{2}$ Layers.
  \emph{Phys. Rev. Lett.} \textbf{1972}, \emph{28}, 299--301\relax
\mciteBstWouldAddEndPuncttrue
\mciteSetBstMidEndSepPunct{\mcitedefaultmidpunct}
{\mcitedefaultendpunct}{\mcitedefaultseppunct}\relax
\EndOfBibitem
\bibitem[Ugeda \latin{et~al.}(2016)Ugeda, Bradley, Zhang, Onishi, Chen, Ruan,
  Ojeda-Aristizabal, Ryu, Edmonds, Tsai, Riss, Mo, Lee, Zettl, Hussain, Shen,
  and Crommie]{nbse2-1}
Ugeda,~M.~M. \latin{et~al.}  Characterization of collective ground states in
  single-layer NbSe2. \emph{Nature Physics} \textbf{2016}, \emph{12},
  92--97\relax
\mciteBstWouldAddEndPuncttrue
\mciteSetBstMidEndSepPunct{\mcitedefaultmidpunct}
{\mcitedefaultendpunct}{\mcitedefaultseppunct}\relax
\EndOfBibitem
\bibitem[Tsen \latin{et~al.}(2016)Tsen, Hunt, Kim, Yuan, Jia, Cava, Hone, Kim,
  Dean, and Pasupathy]{nbse2-2}
Tsen,~A.~W.; Hunt,~B.; Kim,~Y.~D.; Yuan,~Z.~J.; Jia,~S.; Cava,~R.~J.; Hone,~J.;
  Kim,~P.; Dean,~C.~R.; Pasupathy,~A.~N. Nature of the quantum metal in a
  two-dimensional crystalline superconductor. \emph{Nature Physics}
  \textbf{2016}, \emph{12}, 208--212\relax
\mciteBstWouldAddEndPuncttrue
\mciteSetBstMidEndSepPunct{\mcitedefaultmidpunct}
{\mcitedefaultendpunct}{\mcitedefaultseppunct}\relax
\EndOfBibitem
\bibitem[Xi \latin{et~al.}(2016)Xi, Wang, Zhao, Park, Law, Berger, Forr{\'o},
  Shan, and Mak]{nbse2-3}
Xi,~X.; Wang,~Z.; Zhao,~W.; Park,~J.-H.; Law,~K.~T.; Berger,~H.; Forr{\'o},~L.;
  Shan,~J.; Mak,~K.~F. Ising pairing in superconducting NbSe2 atomic layers.
  \emph{Nature Physics} \textbf{2016}, \emph{12}, 139--143\relax
\mciteBstWouldAddEndPuncttrue
\mciteSetBstMidEndSepPunct{\mcitedefaultmidpunct}
{\mcitedefaultendpunct}{\mcitedefaultseppunct}\relax
\EndOfBibitem
\bibitem[Guillam\'on \latin{et~al.}(2008)Guillam\'on, Suderow, Vieira, Cario,
  Diener, and Rodi\`ere]{PhysRevLett.101.166407}
Guillam\'on,~I.; Suderow,~H.; Vieira,~S.; Cario,~L.; Diener,~P.; Rodi\`ere,~P.
  Superconducting Density of States and Vortex Cores of
  2H-${\mathrm{NbS}}_{2}$. \emph{Phys. Rev. Lett.} \textbf{2008}, \emph{101},
  166407\relax
\mciteBstWouldAddEndPuncttrue
\mciteSetBstMidEndSepPunct{\mcitedefaultmidpunct}
{\mcitedefaultendpunct}{\mcitedefaultseppunct}\relax
\EndOfBibitem
\bibitem[Yan \latin{et~al.}(2019)Yan, Khalsa, Schaefer, Jarjour, Rouvimov,
  Nowack, Xing, and Jena]{Yan_2019}
Yan,~R.; Khalsa,~G.; Schaefer,~B.~T.; Jarjour,~A.; Rouvimov,~S.; Nowack,~K.~C.;
  Xing,~H.~G.; Jena,~D. Thickness dependence of superconductivity in ultrathin
  NbS$_2$. \emph{Applied Physics Express} \textbf{2019}, \emph{12},
  023008\relax
\mciteBstWouldAddEndPuncttrue
\mciteSetBstMidEndSepPunct{\mcitedefaultmidpunct}
{\mcitedefaultendpunct}{\mcitedefaultseppunct}\relax
\EndOfBibitem
\bibitem[Li \latin{et~al.}(2016)Li, O'Farrell, Loh, Eda, {\"O}zyilmaz, and
  Castro~Neto]{tise2}
Li,~L.~J.; O'Farrell,~E. C.~T.; Loh,~K.~P.; Eda,~G.; {\"O}zyilmaz,~B.;
  Castro~Neto,~A.~H. Controlling many-body states by the electric-field effect
  in a two-dimensional material. \emph{Nature} \textbf{2016}, \emph{529},
  185--189\relax
\mciteBstWouldAddEndPuncttrue
\mciteSetBstMidEndSepPunct{\mcitedefaultmidpunct}
{\mcitedefaultendpunct}{\mcitedefaultseppunct}\relax
\EndOfBibitem
\bibitem[Navarro-Moratalla \latin{et~al.}(2016)Navarro-Moratalla, Island,
  Ma{\~n}as-Valero, Pinilla-Cienfuegos, Castellanos-Gomez, Quereda,
  Rubio-Bollinger, Chirolli, Silva-Guill{\'e}n, Agra{\"\i}t, Steele, Guinea,
  van~der Zant, and Coronado]{tas2}
Navarro-Moratalla,~E.; Island,~J.~O.; Ma{\~n}as-Valero,~S.;
  Pinilla-Cienfuegos,~E.; Castellanos-Gomez,~A.; Quereda,~J.;
  Rubio-Bollinger,~G.; Chirolli,~L.; Silva-Guill{\'e}n,~J.~A.; Agra{\"\i}t,~N.;
  Steele,~G.~A.; Guinea,~F.; van~der Zant,~H. S.~J.; Coronado,~E. Enhanced
  superconductivity in atomically thin TaS2. \emph{Nature Communications}
  \textbf{2016}, \emph{7}, 11043\relax
\mciteBstWouldAddEndPuncttrue
\mciteSetBstMidEndSepPunct{\mcitedefaultmidpunct}
{\mcitedefaultendpunct}{\mcitedefaultseppunct}\relax
\EndOfBibitem
\bibitem[Lu \latin{et~al.}(2015)Lu, Zheliuk, Leermakers, Yuan, Zeitler, Law,
  and Ye]{doi:10.1126/science.aab2277}
Lu,~J.~M.; Zheliuk,~O.; Leermakers,~I.; Yuan,~N. F.~Q.; Zeitler,~U.;
  Law,~K.~T.; Ye,~J.~T. Evidence for two-dimensional Ising superconductivity in
  gated MoS$_2$. \emph{Science} \textbf{2015}, \emph{350}, 1353--1357\relax
\mciteBstWouldAddEndPuncttrue
\mciteSetBstMidEndSepPunct{\mcitedefaultmidpunct}
{\mcitedefaultendpunct}{\mcitedefaultseppunct}\relax
\EndOfBibitem
\bibitem[Ye \latin{et~al.}(2012)Ye, Zhang, Akashi, Bahramy, Arita, and
  Iwasa]{doi:10.1126/science.1228006}
Ye,~J.~T.; Zhang,~Y.~J.; Akashi,~R.; Bahramy,~M.~S.; Arita,~R.; Iwasa,~Y.
  Superconducting Dome in a Gate-Tuned Band Insulator. \emph{Science}
  \textbf{2012}, \emph{338}, 1193--1196\relax
\mciteBstWouldAddEndPuncttrue
\mciteSetBstMidEndSepPunct{\mcitedefaultmidpunct}
{\mcitedefaultendpunct}{\mcitedefaultseppunct}\relax
\EndOfBibitem
\bibitem[Fu \latin{et~al.}(2017)Fu, Liu, Yuan, Tang, Lian, Xu, Zeng, Chen,
  Wang, Zhou, Xu, Gao, Pan, Wang, Wang, Zhang, Cui, Hwang, and Miao]{mos2-1}
Fu,~Y. \latin{et~al.}  Gated tuned superconductivity and phonon softening in
  monolayer and bilayer MoS2. \emph{npj Quantum Materials} \textbf{2017},
  \emph{2}, 52\relax
\mciteBstWouldAddEndPuncttrue
\mciteSetBstMidEndSepPunct{\mcitedefaultmidpunct}
{\mcitedefaultendpunct}{\mcitedefaultseppunct}\relax
\EndOfBibitem
\bibitem[Costanzo \latin{et~al.}(2016)Costanzo, Jo, Berger, and
  Morpurgo]{mos2-2}
Costanzo,~D.; Jo,~S.; Berger,~H.; Morpurgo,~A.~F. Gate-induced
  superconductivity in atomically thin MoS2 crystals. \emph{Nature
  Nanotechnology} \textbf{2016}, \emph{11}, 339--344\relax
\mciteBstWouldAddEndPuncttrue
\mciteSetBstMidEndSepPunct{\mcitedefaultmidpunct}
{\mcitedefaultendpunct}{\mcitedefaultseppunct}\relax
\EndOfBibitem
\bibitem[Shao \latin{et~al.}(2014)Shao, Lu, Lv, and Sun]{Shao_2014}
Shao,~D.~F.; Lu,~W.~J.; Lv,~H.~Y.; Sun,~Y.~P. Electron-doped phosphorene: A
  potential monolayer superconductor. \emph{{EPL} (Europhysics Letters)}
  \textbf{2014}, \emph{108}, 67004\relax
\mciteBstWouldAddEndPuncttrue
\mciteSetBstMidEndSepPunct{\mcitedefaultmidpunct}
{\mcitedefaultendpunct}{\mcitedefaultseppunct}\relax
\EndOfBibitem
\bibitem[Huang \latin{et~al.}(2015)Huang, Xing, and
  Xing]{doi:10.1063/1.4916100}
Huang,~G.~Q.; Xing,~Z.~W.; Xing,~D.~Y. Prediction of superconductivity in
  Li-intercalated bilayer phosphorene. \emph{Applied Physics Letters}
  \textbf{2015}, \emph{106}, 113107\relax
\mciteBstWouldAddEndPuncttrue
\mciteSetBstMidEndSepPunct{\mcitedefaultmidpunct}
{\mcitedefaultendpunct}{\mcitedefaultseppunct}\relax
\EndOfBibitem
\bibitem[Huang \latin{et~al.}(2016)Huang, Xing, and Xing]{PhysRevB.93.104511}
Huang,~G.~Q.; Xing,~Z.~W.; Xing,~D.~Y. Dynamical stability and
  superconductivity of Li-intercalated bilayer ${\mathrm{MoS}}_{2}$: A
  first-principles prediction. \emph{Phys. Rev. B} \textbf{2016}, \emph{93},
  104511\relax
\mciteBstWouldAddEndPuncttrue
\mciteSetBstMidEndSepPunct{\mcitedefaultmidpunct}
{\mcitedefaultendpunct}{\mcitedefaultseppunct}\relax
\EndOfBibitem
\bibitem[Lugovskoi \latin{et~al.}(2019)Lugovskoi, Katsnelson, and
  Rudenko]{PhysRevB.99.064513}
Lugovskoi,~A.~V.; Katsnelson,~M.~I.; Rudenko,~A.~N. Electron-phonon properties,
  structural stability, and superconductivity of doped antimonene. \emph{Phys.
  Rev. B} \textbf{2019}, \emph{99}, 064513\relax
\mciteBstWouldAddEndPuncttrue
\mciteSetBstMidEndSepPunct{\mcitedefaultmidpunct}
{\mcitedefaultendpunct}{\mcitedefaultseppunct}\relax
\EndOfBibitem
\bibitem[Zhang \latin{et~al.}(2017)Zhang, Zhou, Cui, Zhao, and Liu]{mof}
Zhang,~X.; Zhou,~Y.; Cui,~B.; Zhao,~M.; Liu,~F. Theoretical Discovery of a
  Superconducting Two-Dimensional Metal--Organic Framework. \emph{Nano Letters}
  \textbf{2017}, \emph{17}, 6166--6170\relax
\mciteBstWouldAddEndPuncttrue
\mciteSetBstMidEndSepPunct{\mcitedefaultmidpunct}
{\mcitedefaultendpunct}{\mcitedefaultseppunct}\relax
\EndOfBibitem
\bibitem[Gao \latin{et~al.}(2017)Gao, Li, Yan, and Wang]{PhysRevB.95.024505}
Gao,~M.; Li,~Q.-Z.; Yan,~X.-W.; Wang,~J. Prediction of phonon-mediated
  superconductivity in borophene. \emph{Phys. Rev. B} \textbf{2017}, \emph{95},
  024505\relax
\mciteBstWouldAddEndPuncttrue
\mciteSetBstMidEndSepPunct{\mcitedefaultmidpunct}
{\mcitedefaultendpunct}{\mcitedefaultseppunct}\relax
\EndOfBibitem
\bibitem[Penev \latin{et~al.}(2016)Penev, Kutana, and Yakobson]{boro}
Penev,~E.~S.; Kutana,~A.; Yakobson,~B.~I. Can Two-Dimensional Boron
  Superconduct? \emph{Nano Letters} \textbf{2016}, \emph{16}, 2522--2526\relax
\mciteBstWouldAddEndPuncttrue
\mciteSetBstMidEndSepPunct{\mcitedefaultmidpunct}
{\mcitedefaultendpunct}{\mcitedefaultseppunct}\relax
\EndOfBibitem
\bibitem[Mannix \latin{et~al.}(2015)Mannix, Zhou, Kiraly, Wood, Alducin, Myers,
  Liu, Fisher, Santiago, Guest, Yacaman, Ponce, Oganov, Hersam, and
  Guisinger]{doi:10.1126/science.aad1080}
Mannix,~A.~J.; Zhou,~X.-F.; Kiraly,~B.; Wood,~J.~D.; Alducin,~D.; Myers,~B.~D.;
  Liu,~X.; Fisher,~B.~L.; Santiago,~U.; Guest,~J.~R.; Yacaman,~M.~J.;
  Ponce,~A.; Oganov,~A.~R.; Hersam,~M.~C.; Guisinger,~N.~P. Synthesis of
  borophenes: Anisotropic, two-dimensional boron polymorphs. \emph{Science}
  \textbf{2015}, \emph{350}, 1513--1516\relax
\mciteBstWouldAddEndPuncttrue
\mciteSetBstMidEndSepPunct{\mcitedefaultmidpunct}
{\mcitedefaultendpunct}{\mcitedefaultseppunct}\relax
\EndOfBibitem
\bibitem[Xiao \latin{et~al.}(2016)Xiao, Shao, Lu, Lv, Li, and
  Sun]{doi:10.1063/1.4963179}
Xiao,~R.~C.; Shao,~D.~F.; Lu,~W.~J.; Lv,~H.~Y.; Li,~J.~Y.; Sun,~Y.~P. Enhanced
  superconductivity by strain and carrier-doping in borophene: A first
  principles prediction. \emph{Applied Physics Letters} \textbf{2016},
  \emph{109}, 122604\relax
\mciteBstWouldAddEndPuncttrue
\mciteSetBstMidEndSepPunct{\mcitedefaultmidpunct}
{\mcitedefaultendpunct}{\mcitedefaultseppunct}\relax
\EndOfBibitem
\bibitem[Feng \latin{et~al.}(2016)Feng, Zhang, Zhong, Li, Li, Li, Cheng, Meng,
  Chen, and Wu]{boro2}
Feng,~B.; Zhang,~J.; Zhong,~Q.; Li,~W.; Li,~S.; Li,~H.; Cheng,~P.; Meng,~S.;
  Chen,~L.; Wu,~K. Experimental realization of two-dimensional boron sheets.
  \emph{Nature Chemistry} \textbf{2016}, \emph{8}, 563--568\relax
\mciteBstWouldAddEndPuncttrue
\mciteSetBstMidEndSepPunct{\mcitedefaultmidpunct}
{\mcitedefaultendpunct}{\mcitedefaultseppunct}\relax
\EndOfBibitem
\bibitem[Yan \latin{et~al.}(2020)Yan, Liu, Li, Tang, He, Huang, Wang, and
  Zhou]{b2o}
Yan,~L.; Liu,~P.-F.; Li,~H.; Tang,~Y.; He,~J.; Huang,~X.; Wang,~B.-T.; Zhou,~L.
  Theoretical dissection of superconductivity in two-dimensional honeycomb
  borophene oxide B2O crystal with a high stability. \emph{npj Computational
  Materials} \textbf{2020}, \emph{6}, 94\relax
\mciteBstWouldAddEndPuncttrue
\mciteSetBstMidEndSepPunct{\mcitedefaultmidpunct}
{\mcitedefaultendpunct}{\mcitedefaultseppunct}\relax
\EndOfBibitem
\bibitem[Dai \latin{et~al.}(2012)Dai, Li, Yang, and Hou]{C2NR12018F}
Dai,~J.; Li,~Z.; Yang,~J.; Hou,~J. A first-principles prediction of
  two-dimensional superconductivity in pristine B2C single layers.
  \emph{Nanoscale} \textbf{2012}, \emph{4}, 3032--3035\relax
\mciteBstWouldAddEndPuncttrue
\mciteSetBstMidEndSepPunct{\mcitedefaultmidpunct}
{\mcitedefaultendpunct}{\mcitedefaultseppunct}\relax
\EndOfBibitem
\bibitem[Modak \latin{et~al.}(2021)Modak, Verma, and
  Mishra]{PhysRevB.104.054504}
Modak,~P.; Verma,~A.~K.; Mishra,~A.~K. Prediction of superconductivity at 70 K
  in a pristine monolayer of LiBC. \emph{Phys. Rev. B} \textbf{2021},
  \emph{104}, 054504\relax
\mciteBstWouldAddEndPuncttrue
\mciteSetBstMidEndSepPunct{\mcitedefaultmidpunct}
{\mcitedefaultendpunct}{\mcitedefaultseppunct}\relax
\EndOfBibitem
\bibitem[Bekaert \latin{et~al.}(2017)Bekaert, Aperis, Partoens, Oppeneer, and
  Milo\ifmmode \check{s}\else \v{s}\fi{}evi\ifmmode~\acute{c}\else
  \'{c}\fi{}]{PhysRevB.96.094510}
Bekaert,~J.; Aperis,~A.; Partoens,~B.; Oppeneer,~P.~M.; Milo\ifmmode
  \check{s}\else \v{s}\fi{}evi\ifmmode~\acute{c}\else \'{c}\fi{},~M.~V.
  Evolution of multigap superconductivity in the atomically thin limit:
  Strain-enhanced three-gap superconductivity in monolayer
  ${\mathrm{MgB}}_{2}$. \emph{Phys. Rev. B} \textbf{2017}, \emph{96},
  094510\relax
\mciteBstWouldAddEndPuncttrue
\mciteSetBstMidEndSepPunct{\mcitedefaultmidpunct}
{\mcitedefaultendpunct}{\mcitedefaultseppunct}\relax
\EndOfBibitem
\bibitem[Lei \latin{et~al.}(2017)Lei, Kutana, and Yakobson]{C7TC00789B}
Lei,~J.; Kutana,~A.; Yakobson,~B.~I. Predicting stable phase monolayer Mo2C
  (MXene){,} a superconductor with chemically-tunable critical temperature.
  \emph{J. Mater. Chem. C} \textbf{2017}, \emph{5}, 3438--3444\relax
\mciteBstWouldAddEndPuncttrue
\mciteSetBstMidEndSepPunct{\mcitedefaultmidpunct}
{\mcitedefaultendpunct}{\mcitedefaultseppunct}\relax
\EndOfBibitem
\bibitem[Singh \latin{et~al.}(2022)Singh, Romero, Mella, Eremeev, Mu{\~n}oz,
  Alexandrova, Rabe, Vanderbilt, and Mu{\~n}oz]{Mg2B4C2}
Singh,~S.; Romero,~A.~H.; Mella,~J.; Eremeev,~V.; Mu{\~n}oz,~E.;
  Alexandrova,~A.~N.; Rabe,~K.~M.; Vanderbilt,~D.; Mu{\~n}oz,~F.
  High-temperature phonon-mediated superconductivity in monolayer Mg2B4C2.
  \emph{npj Quantum Materials} \textbf{2022}, \emph{7}, 37\relax
\mciteBstWouldAddEndPuncttrue
\mciteSetBstMidEndSepPunct{\mcitedefaultmidpunct}
{\mcitedefaultendpunct}{\mcitedefaultseppunct}\relax
\EndOfBibitem
\bibitem[Campi \latin{et~al.}(2021)Campi, Kumari, and
  Marzari]{doi:10.1021/acs.nanolett.0c05125}
Campi,~D.; Kumari,~S.; Marzari,~N. Prediction of Phonon-Mediated
  Superconductivity with High Critical Temperature in the Two-Dimensional
  Topological Semimetal W2N3. \emph{Nano Letters} \textbf{2021}, \emph{21},
  3435--3442, PMID: 33856216\relax
\mciteBstWouldAddEndPuncttrue
\mciteSetBstMidEndSepPunct{\mcitedefaultmidpunct}
{\mcitedefaultendpunct}{\mcitedefaultseppunct}\relax
\EndOfBibitem
\bibitem[Roberts(1976)]{roberts1976survey}
Roberts,~B.~W. Survey of superconductive materials and critical evaluation of
  selected properties. \emph{Journal of Physical and Chemical Reference Data}
  \textbf{1976}, \emph{5}, 581--822\relax
\mciteBstWouldAddEndPuncttrue
\mciteSetBstMidEndSepPunct{\mcitedefaultmidpunct}
{\mcitedefaultendpunct}{\mcitedefaultseppunct}\relax
\EndOfBibitem
\bibitem[Kihlstrom \latin{et~al.}(1985)Kihlstrom, Simon, and
  Wolf]{kihlstrom1985tunneling}
Kihlstrom,~K.; Simon,~R.; Wolf,~S. Tunneling $\alpha$2F ($\omega$) on high Tc
  A15 and B1 compounds. \emph{Physica B+ C} \textbf{1985}, \emph{135},
  198--202\relax
\mciteBstWouldAddEndPuncttrue
\mciteSetBstMidEndSepPunct{\mcitedefaultmidpunct}
{\mcitedefaultendpunct}{\mcitedefaultseppunct}\relax
\EndOfBibitem
\bibitem[Stewart(2015)]{stewart2015superconductivity}
Stewart,~G.~R. Superconductivity in the A15 structure. \emph{Physica C:
  Superconductivity and its Applications} \textbf{2015}, \emph{514},
  28--35\relax
\mciteBstWouldAddEndPuncttrue
\mciteSetBstMidEndSepPunct{\mcitedefaultmidpunct}
{\mcitedefaultendpunct}{\mcitedefaultseppunct}\relax
\EndOfBibitem
\bibitem[Ivanovskii(2003)]{ivanovskii2003band}
Ivanovskii,~A. Band structure and properties of superconducting MgB2 and
  related compounds (a review). \emph{Physics of the Solid State}
  \textbf{2003}, \emph{45}, 1829--1859\relax
\mciteBstWouldAddEndPuncttrue
\mciteSetBstMidEndSepPunct{\mcitedefaultmidpunct}
{\mcitedefaultendpunct}{\mcitedefaultseppunct}\relax
\EndOfBibitem
\bibitem[Buzea and Yamashita(2001)Buzea, and Yamashita]{buzea2001review}
Buzea,~C.; Yamashita,~T. Review of the superconducting properties of MgB2.
  \emph{Superconductor Science and Technology} \textbf{2001}, \emph{14},
  R115\relax
\mciteBstWouldAddEndPuncttrue
\mciteSetBstMidEndSepPunct{\mcitedefaultmidpunct}
{\mcitedefaultendpunct}{\mcitedefaultseppunct}\relax
\EndOfBibitem
\bibitem[Nagamatsu \latin{et~al.}(2001)Nagamatsu, Nakagawa, Muranaka, Zenitani,
  and Akimitsu]{nagamatsu2001superconductivity}
Nagamatsu,~J.; Nakagawa,~N.; Muranaka,~T.; Zenitani,~Y.; Akimitsu,~J.
  Superconductivity at 39 K in magnesium diboride. \emph{nature} \textbf{2001},
  \emph{410}, 63--64\relax
\mciteBstWouldAddEndPuncttrue
\mciteSetBstMidEndSepPunct{\mcitedefaultmidpunct}
{\mcitedefaultendpunct}{\mcitedefaultseppunct}\relax
\EndOfBibitem
\bibitem[Plakida(2010)]{plakida2010high}
Plakida,~N. \emph{High-temperature cuprate superconductors: Experiment, theory,
  and applications}; Springer Science \& Business Media, 2010; Vol. 166\relax
\mciteBstWouldAddEndPuncttrue
\mciteSetBstMidEndSepPunct{\mcitedefaultmidpunct}
{\mcitedefaultendpunct}{\mcitedefaultseppunct}\relax
\EndOfBibitem
\bibitem[Hosono and Kuroki(2015)Hosono, and Kuroki]{hosono2015iron}
Hosono,~H.; Kuroki,~K. Iron-based superconductors: Current status of materials
  and pairing mechanism. \emph{Physica C: Superconductivity and its
  Applications} \textbf{2015}, \emph{514}, 399--422\relax
\mciteBstWouldAddEndPuncttrue
\mciteSetBstMidEndSepPunct{\mcitedefaultmidpunct}
{\mcitedefaultendpunct}{\mcitedefaultseppunct}\relax
\EndOfBibitem
\bibitem[Shipley \latin{et~al.}(2020)Shipley, Hutcheon, Johnson, Needs, and
  Pickard]{shipley2020stability}
Shipley,~A.~M.; Hutcheon,~M.~J.; Johnson,~M.~S.; Needs,~R.~J.; Pickard,~C.~J.
  Stability and superconductivity of lanthanum and yttrium decahydrides.
  \emph{Physical Review B} \textbf{2020}, \emph{101}, 224511\relax
\mciteBstWouldAddEndPuncttrue
\mciteSetBstMidEndSepPunct{\mcitedefaultmidpunct}
{\mcitedefaultendpunct}{\mcitedefaultseppunct}\relax
\EndOfBibitem
\bibitem[Liu \latin{et~al.}(2017)Liu, Naumov, Hoffmann, Ashcroft, and
  Hemley]{liu2017potential}
Liu,~H.; Naumov,~I.~I.; Hoffmann,~R.; Ashcroft,~N.; Hemley,~R.~J. Potential
  high-Tc superconducting lanthanum and yttrium hydrides at high pressure.
  \emph{Proceedings of the National Academy of Sciences} \textbf{2017},
  \emph{114}, 6990--6995\relax
\mciteBstWouldAddEndPuncttrue
\mciteSetBstMidEndSepPunct{\mcitedefaultmidpunct}
{\mcitedefaultendpunct}{\mcitedefaultseppunct}\relax
\EndOfBibitem
\bibitem[Drozdov \latin{et~al.}(2015)Drozdov, Eremets, Troyan, Ksenofontov, and
  Shylin]{drozdov2015conventional}
Drozdov,~A.; Eremets,~M.; Troyan,~I.; Ksenofontov,~V.; Shylin,~S.~I.
  Conventional superconductivity at 203 kelvin at high pressures in the sulfur
  hydride system. \emph{Nature} \textbf{2015}, \emph{525}, 73--76\relax
\mciteBstWouldAddEndPuncttrue
\mciteSetBstMidEndSepPunct{\mcitedefaultmidpunct}
{\mcitedefaultendpunct}{\mcitedefaultseppunct}\relax
\EndOfBibitem
\bibitem[Poole \latin{et~al.}(2013)Poole, Farach, and
  Creswick]{poole2013superconductivity}
Poole,~C.~P.; Farach,~H.~A.; Creswick,~R.~J. \emph{Superconductivity}; Academic
  press, 2013\relax
\mciteBstWouldAddEndPuncttrue
\mciteSetBstMidEndSepPunct{\mcitedefaultmidpunct}
{\mcitedefaultendpunct}{\mcitedefaultseppunct}\relax
\EndOfBibitem
\bibitem[Shipley \latin{et~al.}(2021)Shipley, Hutcheon, Needs, and
  Pickard]{shipley2021high}
Shipley,~A.~M.; Hutcheon,~M.~J.; Needs,~R.~J.; Pickard,~C.~J. High-throughput
  discovery of high-temperature conventional superconductors. \emph{Physical
  Review B} \textbf{2021}, \emph{104}, 054501\relax
\mciteBstWouldAddEndPuncttrue
\mciteSetBstMidEndSepPunct{\mcitedefaultmidpunct}
{\mcitedefaultendpunct}{\mcitedefaultseppunct}\relax
\EndOfBibitem
\bibitem[Yuan \latin{et~al.}(2019)Yuan, Stanev, Gao, Takeuchi, and
  Jin]{yuan2019recent}
Yuan,~J.; Stanev,~V.; Gao,~C.; Takeuchi,~I.; Jin,~K. Recent advances in
  high-throughput superconductivity research. \emph{Superconductor Science and
  Technology} \textbf{2019}, \emph{32}, 123001\relax
\mciteBstWouldAddEndPuncttrue
\mciteSetBstMidEndSepPunct{\mcitedefaultmidpunct}
{\mcitedefaultendpunct}{\mcitedefaultseppunct}\relax
\EndOfBibitem
\bibitem[Rodriguez \latin{et~al.}(1990)Rodriguez, Liechtenstein, Mazin, Jepsen,
  Andersen, and Methfessel]{rodriguez1990optical}
Rodriguez,~C.; Liechtenstein,~A.; Mazin,~I.; Jepsen,~O.; Andersen,~O.;
  Methfessel,~M. Optical near-zone-center phonons and their interaction with
  electrons in YBa 2 Cu 3 O 7: results of the local-density approximation.
  \emph{Physical Review B} \textbf{1990}, \emph{42}, 2692\relax
\mciteBstWouldAddEndPuncttrue
\mciteSetBstMidEndSepPunct{\mcitedefaultmidpunct}
{\mcitedefaultendpunct}{\mcitedefaultseppunct}\relax
\EndOfBibitem
\bibitem[Subedi \latin{et~al.}(2013)Subedi, Ortenzi, and
  Boeri]{subedi2013electron}
Subedi,~A.; Ortenzi,~L.; Boeri,~L. Electron-phonon superconductivity in A Pt 3
  P (A= Sr, Ca, La) compounds: From weak to strong coupling. \emph{Physical
  Review B} \textbf{2013}, \emph{87}, 144504\relax
\mciteBstWouldAddEndPuncttrue
\mciteSetBstMidEndSepPunct{\mcitedefaultmidpunct}
{\mcitedefaultendpunct}{\mcitedefaultseppunct}\relax
\EndOfBibitem
\bibitem[Duan \latin{et~al.}(2019)Duan, Yu, Xie, and Cui]{duan2019ab}
Duan,~D.; Yu,~H.; Xie,~H.; Cui,~T. Ab initio approach and its impact on
  superconductivity. \emph{Journal of Superconductivity and Novel Magnetism}
  \textbf{2019}, \emph{32}, 53--60\relax
\mciteBstWouldAddEndPuncttrue
\mciteSetBstMidEndSepPunct{\mcitedefaultmidpunct}
{\mcitedefaultendpunct}{\mcitedefaultseppunct}\relax
\EndOfBibitem
\bibitem[Kolmogorov \latin{et~al.}(2010)Kolmogorov, Shah, Margine, Bialon,
  Hammerschmidt, and Drautz]{kolmogorov2010new}
Kolmogorov,~A.; Shah,~S.; Margine,~E.; Bialon,~A.; Hammerschmidt,~T.;
  Drautz,~R. New superconducting and semiconducting Fe-B compounds predicted
  with an ab initio evolutionary search. \emph{Physical review letters}
  \textbf{2010}, \emph{105}, 217003\relax
\mciteBstWouldAddEndPuncttrue
\mciteSetBstMidEndSepPunct{\mcitedefaultmidpunct}
{\mcitedefaultendpunct}{\mcitedefaultseppunct}\relax
\EndOfBibitem
\bibitem[Gao \latin{et~al.}(2010)Gao, Oganov, Li, Li, Wang, Cui, Ma, Bergara,
  Lyakhov, Iitaka, \latin{et~al.} others]{gao2010high}
Gao,~G.; Oganov,~A.~R.; Li,~P.; Li,~Z.; Wang,~H.; Cui,~T.; Ma,~Y.; Bergara,~A.;
  Lyakhov,~A.~O.; Iitaka,~T., \latin{et~al.}  High-pressure crystal structures
  and superconductivity of Stannane (SnH4). \emph{Proceedings of the National
  Academy of Sciences} \textbf{2010}, \emph{107}, 1317--1320\relax
\mciteBstWouldAddEndPuncttrue
\mciteSetBstMidEndSepPunct{\mcitedefaultmidpunct}
{\mcitedefaultendpunct}{\mcitedefaultseppunct}\relax
\EndOfBibitem
\bibitem[Duan \latin{et~al.}(2014)Duan, Liu, Tian, Li, Huang, Zhao, Yu, Liu,
  Tian, and Cui]{duan2014pressure}
Duan,~D.; Liu,~Y.; Tian,~F.; Li,~D.; Huang,~X.; Zhao,~Z.; Yu,~H.; Liu,~B.;
  Tian,~W.; Cui,~T. Pressure-induced metallization of dense (H2S) 2H2 with
  high-Tc superconductivity. \emph{Scientific reports} \textbf{2014}, \emph{4},
  1--6\relax
\mciteBstWouldAddEndPuncttrue
\mciteSetBstMidEndSepPunct{\mcitedefaultmidpunct}
{\mcitedefaultendpunct}{\mcitedefaultseppunct}\relax
\EndOfBibitem
\bibitem[Cooper and Feldman(2010)Cooper, and Feldman]{cooper2010bcs}
Cooper,~L.~N.; Feldman,~D. \emph{BCS: 50 years}; World scientific, 2010\relax
\mciteBstWouldAddEndPuncttrue
\mciteSetBstMidEndSepPunct{\mcitedefaultmidpunct}
{\mcitedefaultendpunct}{\mcitedefaultseppunct}\relax
\EndOfBibitem
\bibitem[Giustino(2017)]{giustino2017electron}
Giustino,~F. Electron-phonon interactions from first principles. \emph{Reviews
  of Modern Physics} \textbf{2017}, \emph{89}, 015003\relax
\mciteBstWouldAddEndPuncttrue
\mciteSetBstMidEndSepPunct{\mcitedefaultmidpunct}
{\mcitedefaultendpunct}{\mcitedefaultseppunct}\relax
\EndOfBibitem
\bibitem[Kawamura \latin{et~al.}(2020)Kawamura, Hizume, and
  Ozaki]{PhysRevB.101.134511}
Kawamura,~M.; Hizume,~Y.; Ozaki,~T. Benchmark of density functional theory for
  superconductors in elemental materials. \emph{Phys. Rev. B} \textbf{2020},
  \emph{101}, 134511\relax
\mciteBstWouldAddEndPuncttrue
\mciteSetBstMidEndSepPunct{\mcitedefaultmidpunct}
{\mcitedefaultendpunct}{\mcitedefaultseppunct}\relax
\EndOfBibitem
\bibitem[Choudhary \latin{et~al.}(2020)Choudhary, Garrity, Reid, DeCost,
  Biacchi, Hight~Walker, Trautt, Hattrick-Simpers, Kusne, Centrone,
  \latin{et~al.} others]{choudhary2020joint}
Choudhary,~K.; Garrity,~K.~F.; Reid,~A.~C.; DeCost,~B.; Biacchi,~A.~J.;
  Hight~Walker,~A.~R.; Trautt,~Z.; Hattrick-Simpers,~J.; Kusne,~A.~G.;
  Centrone,~A., \latin{et~al.}  The joint automated repository for various
  integrated simulations (JARVIS) for data-driven materials design. \emph{npj
  Computational Materials} \textbf{2020}, \emph{6}, 1--13\relax
\mciteBstWouldAddEndPuncttrue
\mciteSetBstMidEndSepPunct{\mcitedefaultmidpunct}
{\mcitedefaultendpunct}{\mcitedefaultseppunct}\relax
\EndOfBibitem
\bibitem[Allen and Dynes(1975)Allen, and Dynes]{allen1975transition}
Allen,~P.~B.; Dynes,~R. Transition temperature of strong-coupled
  superconductors reanalyzed. \emph{Physical Review B} \textbf{1975},
  \emph{12}, 905\relax
\mciteBstWouldAddEndPuncttrue
\mciteSetBstMidEndSepPunct{\mcitedefaultmidpunct}
{\mcitedefaultendpunct}{\mcitedefaultseppunct}\relax
\EndOfBibitem
\bibitem[Choudhary \latin{et~al.}(2018)Choudhary, Zhang, Reid, Chowdhury,
  Van~Nguyen, Trautt, Newrock, Congo, and Tavazza]{choudhary2018computational}
Choudhary,~K.; Zhang,~Q.; Reid,~A.~C.; Chowdhury,~S.; Van~Nguyen,~N.;
  Trautt,~Z.; Newrock,~M.~W.; Congo,~F.~Y.; Tavazza,~F. Computational screening
  of high-performance optoelectronic materials using OptB88vdW and TB-mBJ
  formalisms. \emph{Scientific data} \textbf{2018}, \emph{5}, 1--12\relax
\mciteBstWouldAddEndPuncttrue
\mciteSetBstMidEndSepPunct{\mcitedefaultmidpunct}
{\mcitedefaultendpunct}{\mcitedefaultseppunct}\relax
\EndOfBibitem
\bibitem[Choudhary \latin{et~al.}(2017)Choudhary, Kalish, Beams, and
  Tavazza]{choudhary2017high}
Choudhary,~K.; Kalish,~I.; Beams,~R.; Tavazza,~F. High-throughput
  identification and characterization of two-dimensional materials using
  density functional theory. \emph{Scientific reports} \textbf{2017}, \emph{7},
  1--16\relax
\mciteBstWouldAddEndPuncttrue
\mciteSetBstMidEndSepPunct{\mcitedefaultmidpunct}
{\mcitedefaultendpunct}{\mcitedefaultseppunct}\relax
\EndOfBibitem
\bibitem[Choudhary \latin{et~al.}(2019)Choudhary, Bercx, Jiang, Pachter,
  Lamoen, and Tavazza]{choudhary2019accelerated}
Choudhary,~K.; Bercx,~M.; Jiang,~J.; Pachter,~R.; Lamoen,~D.; Tavazza,~F.
  Accelerated discovery of efficient solar cell materials using quantum and
  machine-learning methods. \emph{Chemistry of Materials} \textbf{2019},
  \emph{31}, 5900--5908\relax
\mciteBstWouldAddEndPuncttrue
\mciteSetBstMidEndSepPunct{\mcitedefaultmidpunct}
{\mcitedefaultendpunct}{\mcitedefaultseppunct}\relax
\EndOfBibitem
\bibitem[Choudhary \latin{et~al.}(2021)Choudhary, Garrity, Ghimire, Anand, and
  Tavazza]{choudhary2021high}
Choudhary,~K.; Garrity,~K.~F.; Ghimire,~N.~J.; Anand,~N.; Tavazza,~F.
  High-throughput search for magnetic topological materials using spin-orbit
  spillage, machine learning, and experiments. \emph{Physical Review B}
  \textbf{2021}, \emph{103}, 155131\relax
\mciteBstWouldAddEndPuncttrue
\mciteSetBstMidEndSepPunct{\mcitedefaultmidpunct}
{\mcitedefaultendpunct}{\mcitedefaultseppunct}\relax
\EndOfBibitem
\bibitem[Choudhary \latin{et~al.}(2019)Choudhary, Garrity, and
  Tavazza]{choudhary2019high}
Choudhary,~K.; Garrity,~K.~F.; Tavazza,~F. High-throughput discovery of
  topologically non-trivial materials using spin-orbit spillage.
  \emph{Scientific reports} \textbf{2019}, \emph{9}, 1--8\relax
\mciteBstWouldAddEndPuncttrue
\mciteSetBstMidEndSepPunct{\mcitedefaultmidpunct}
{\mcitedefaultendpunct}{\mcitedefaultseppunct}\relax
\EndOfBibitem
\bibitem[Choudhary \latin{et~al.}(2020)Choudhary, Garrity, Jiang, Pachter, and
  Tavazza]{choudhary2020computational}
Choudhary,~K.; Garrity,~K.~F.; Jiang,~J.; Pachter,~R.; Tavazza,~F.
  Computational search for magnetic and non-magnetic 2D topological materials
  using unified spin--orbit spillage screening. \emph{NPJ Computational
  Materials} \textbf{2020}, \emph{6}, 1--8\relax
\mciteBstWouldAddEndPuncttrue
\mciteSetBstMidEndSepPunct{\mcitedefaultmidpunct}
{\mcitedefaultendpunct}{\mcitedefaultseppunct}\relax
\EndOfBibitem
\bibitem[Choudhary \latin{et~al.}(2018)Choudhary, Cheon, Reed, and
  Tavazza]{choudhary2018elastic}
Choudhary,~K.; Cheon,~G.; Reed,~E.; Tavazza,~F. Elastic properties of bulk and
  low-dimensional materials using van der Waals density functional.
  \emph{Physical Review B} \textbf{2018}, \emph{98}, 014107\relax
\mciteBstWouldAddEndPuncttrue
\mciteSetBstMidEndSepPunct{\mcitedefaultmidpunct}
{\mcitedefaultendpunct}{\mcitedefaultseppunct}\relax
\EndOfBibitem
\bibitem[Choudhary \latin{et~al.}(2020)Choudhary, Garrity, Sharma, Biacchi,
  Hight~Walker, and Tavazza]{choudhary2020high}
Choudhary,~K.; Garrity,~K.~F.; Sharma,~V.; Biacchi,~A.~J.; Hight~Walker,~A.~R.;
  Tavazza,~F. High-throughput density functional perturbation theory and
  machine learning predictions of infrared, piezoelectric, and dielectric
  responses. \emph{NPJ Computational Materials} \textbf{2020}, \emph{6},
  1--13\relax
\mciteBstWouldAddEndPuncttrue
\mciteSetBstMidEndSepPunct{\mcitedefaultmidpunct}
{\mcitedefaultendpunct}{\mcitedefaultseppunct}\relax
\EndOfBibitem
\bibitem[Choudhary \latin{et~al.}(2020)Choudhary, Ansari, Mazin, and
  Sauer]{choudhary2020density}
Choudhary,~K.; Ansari,~J.~N.; Mazin,~I.~I.; Sauer,~K.~L. Density functional
  theory-based electric field gradient database. \emph{Scientific Data}
  \textbf{2020}, \emph{7}, 1--10\relax
\mciteBstWouldAddEndPuncttrue
\mciteSetBstMidEndSepPunct{\mcitedefaultmidpunct}
{\mcitedefaultendpunct}{\mcitedefaultseppunct}\relax
\EndOfBibitem
\bibitem[Wines \latin{et~al.}(2023)Wines, Choudhary, and
  Tavazza]{https://doi.org/10.48550/arxiv.2209.10379}
Wines,~D.; Choudhary,~K.; Tavazza,~F. Systematic DFT+U and Quantum Monte Carlo
  Benchmark of Magnetic Two-Dimensional (2D) CrX3 (X = I, Br, Cl, F). \emph{The
  Journal of Physical Chemistry C} \textbf{2023}, https://doi.org/10.1021/acs.jpcc.2c06733 \relax
\mciteBstWouldAddEndPunctfalse
\mciteSetBstMidEndSepPunct{\mcitedefaultmidpunct}
{}{\mcitedefaultseppunct}\relax
\EndOfBibitem
\bibitem[Choudhary and Garrity(2022)Choudhary, and
  Garrity]{https://doi.org/10.48550/arxiv.2205.00060}
Choudhary,~K.; Garrity,~K. Designing high-TC superconductors with BCS-inspired
  screening, density functional theory, and deep-learning. \emph{npj
  Computational Materials} \textbf{2022}, \emph{8}, 244\relax
\mciteBstWouldAddEndPuncttrue
\mciteSetBstMidEndSepPunct{\mcitedefaultmidpunct}
{\mcitedefaultendpunct}{\mcitedefaultseppunct}\relax
\EndOfBibitem
\bibitem[Choudhary and Tavazza(2019)Choudhary, and
  Tavazza]{choudhary2019convergence}
Choudhary,~K.; Tavazza,~F. Convergence and machine learning predictions of
  Monkhorst-Pack k-points and plane-wave cut-off in high-throughput DFT
  calculations. \emph{Computational materials science} \textbf{2019},
  \emph{161}, 300--308\relax
\mciteBstWouldAddEndPuncttrue
\mciteSetBstMidEndSepPunct{\mcitedefaultmidpunct}
{\mcitedefaultendpunct}{\mcitedefaultseppunct}\relax
\EndOfBibitem
\bibitem[Bardeen \latin{et~al.}(1957)Bardeen, Cooper, and
  Schrieffer]{bardeen1957theory}
Bardeen,~J.; Cooper,~L.~N.; Schrieffer,~J.~R. Theory of superconductivity.
  \emph{Physical review} \textbf{1957}, \emph{108}, 1175\relax
\mciteBstWouldAddEndPuncttrue
\mciteSetBstMidEndSepPunct{\mcitedefaultmidpunct}
{\mcitedefaultendpunct}{\mcitedefaultseppunct}\relax
\EndOfBibitem
\bibitem[Anderson(1963)]{anderson1963simplified}
Anderson,~O.~L. A simplified method for calculating the Debye temperature from
  elastic constants. \emph{Journal of Physics and Chemistry of Solids}
  \textbf{1963}, \emph{24}, 909--917\relax
\mciteBstWouldAddEndPuncttrue
\mciteSetBstMidEndSepPunct{\mcitedefaultmidpunct}
{\mcitedefaultendpunct}{\mcitedefaultseppunct}\relax
\EndOfBibitem
\bibitem[Kresse and Furthm{\"u}ller(1996)Kresse, and
  Furthm{\"u}ller]{kresse1996efficient}
Kresse,~G.; Furthm{\"u}ller,~J. Efficient iterative schemes for ab initio
  total-energy calculations using a plane-wave basis set. \emph{Physical review
  B} \textbf{1996}, \emph{54}, 11169\relax
\mciteBstWouldAddEndPuncttrue
\mciteSetBstMidEndSepPunct{\mcitedefaultmidpunct}
{\mcitedefaultendpunct}{\mcitedefaultseppunct}\relax
\EndOfBibitem
\bibitem[Kresse and Furthm{\"u}ller(1996)Kresse, and
  Furthm{\"u}ller]{kresse1996efficiency}
Kresse,~G.; Furthm{\"u}ller,~J. Efficiency of ab-initio total energy
  calculations for metals and semiconductors using a plane-wave basis set.
  \emph{Computational materials science} \textbf{1996}, \emph{6}, 15--50\relax
\mciteBstWouldAddEndPuncttrue
\mciteSetBstMidEndSepPunct{\mcitedefaultmidpunct}
{\mcitedefaultendpunct}{\mcitedefaultseppunct}\relax
\EndOfBibitem
\bibitem[Klime{\v{s}} \latin{et~al.}(2009)Klime{\v{s}}, Bowler, and
  Michaelides]{klimevs2009chemical}
Klime{\v{s}},~J.; Bowler,~D.~R.; Michaelides,~A. Chemical accuracy for the van
  der Waals density functional. \emph{Journal of Physics: Condensed Matter}
  \textbf{2009}, \emph{22}, 022201\relax
\mciteBstWouldAddEndPuncttrue
\mciteSetBstMidEndSepPunct{\mcitedefaultmidpunct}
{\mcitedefaultendpunct}{\mcitedefaultseppunct}\relax
\EndOfBibitem
\bibitem[Giannozzi \latin{et~al.}(2009)Giannozzi, Baroni, Bonini, Calandra,
  Car, Cavazzoni, Ceresoli, Chiarotti, Cococcioni, Dabo, \latin{et~al.}
  others]{giannozzi2009quantum}
Giannozzi,~P.; Baroni,~S.; Bonini,~N.; Calandra,~M.; Car,~R.; Cavazzoni,~C.;
  Ceresoli,~D.; Chiarotti,~G.~L.; Cococcioni,~M.; Dabo,~I., \latin{et~al.}
  QUANTUM ESPRESSO: a modular and open-source software project for quantum
  simulations of materials. \emph{Journal of physics: Condensed matter}
  \textbf{2009}, \emph{21}, 395502\relax
\mciteBstWouldAddEndPuncttrue
\mciteSetBstMidEndSepPunct{\mcitedefaultmidpunct}
{\mcitedefaultendpunct}{\mcitedefaultseppunct}\relax
\EndOfBibitem
\bibitem[Giannozzi \latin{et~al.}(2020)Giannozzi, Baseggio, Bonf{\`a}, Brunato,
  Car, Carnimeo, Cavazzoni, De~Gironcoli, Delugas, Ferrari~Ruffino,
  \latin{et~al.} others]{giannozzi2020quantum}
Giannozzi,~P.; Baseggio,~O.; Bonf{\`a},~P.; Brunato,~D.; Car,~R.; Carnimeo,~I.;
  Cavazzoni,~C.; De~Gironcoli,~S.; Delugas,~P.; Ferrari~Ruffino,~F.,
  \latin{et~al.}  Quantum ESPRESSO toward the exascale. \emph{The Journal of
  Chemical Physics} \textbf{2020}, \emph{152}, 154105\relax
\mciteBstWouldAddEndPuncttrue
\mciteSetBstMidEndSepPunct{\mcitedefaultmidpunct}
{\mcitedefaultendpunct}{\mcitedefaultseppunct}\relax
\EndOfBibitem
\bibitem[Wierzbowska \latin{et~al.}(2005)Wierzbowska, de~Gironcoli, and
  Giannozzi]{wierzbowska2005origins}
Wierzbowska,~M.; de~Gironcoli,~S.; Giannozzi,~P. Origins of low-and
  high-pressure discontinuities of $ Tc $ in niobium. \textbf{2005}, arXiv:cond-mat/0504077. \emph{arXiv
  preprint cond-mat/0504077}. https://doi.org/10.48550/arXiv.cond-mat/0504077 (accessed October 1, 2022) \relax
\mciteBstWouldAddEndPunctfalse
\mciteSetBstMidEndSepPunct{\mcitedefaultmidpunct}
{}{\mcitedefaultseppunct}\relax
\EndOfBibitem
\bibitem[Kawamura \latin{et~al.}(2014)Kawamura, Gohda, and
  Tsuneyuki]{kawamura2014improved}
Kawamura,~M.; Gohda,~Y.; Tsuneyuki,~S. Improved tetrahedron method for the
  Brillouin-zone integration applicable to response functions. \emph{Physical
  Review B} \textbf{2014}, \emph{89}, 094515\relax
\mciteBstWouldAddEndPuncttrue
\mciteSetBstMidEndSepPunct{\mcitedefaultmidpunct}
{\mcitedefaultendpunct}{\mcitedefaultseppunct}\relax
\EndOfBibitem
\bibitem[Ponc{\'e} \latin{et~al.}(2016)Ponc{\'e}, Margine, Verdi, and
  Giustino]{ponce2016epw}
Ponc{\'e},~S.; Margine,~E.~R.; Verdi,~C.; Giustino,~F. EPW: Electron--phonon
  coupling, transport and superconducting properties using maximally localized
  Wannier functions. \emph{Computer Physics Communications} \textbf{2016},
  \emph{209}, 116--133\relax
\mciteBstWouldAddEndPuncttrue
\mciteSetBstMidEndSepPunct{\mcitedefaultmidpunct}
{\mcitedefaultendpunct}{\mcitedefaultseppunct}\relax
\EndOfBibitem
\bibitem[Baroni \latin{et~al.}(1987)Baroni, Giannozzi, and
  Testa]{baroni1987green}
Baroni,~S.; Giannozzi,~P.; Testa,~A. Green’s-function approach to linear
  response in solids. \emph{Physical review letters} \textbf{1987}, \emph{58},
  1861\relax
\mciteBstWouldAddEndPuncttrue
\mciteSetBstMidEndSepPunct{\mcitedefaultmidpunct}
{\mcitedefaultendpunct}{\mcitedefaultseppunct}\relax
\EndOfBibitem
\bibitem[Gonze(1995)]{gonze1995perturbation}
Gonze,~X. Perturbation expansion of variational principles at arbitrary order.
  \emph{Physical Review A} \textbf{1995}, \emph{52}, 1086\relax
\mciteBstWouldAddEndPuncttrue
\mciteSetBstMidEndSepPunct{\mcitedefaultmidpunct}
{\mcitedefaultendpunct}{\mcitedefaultseppunct}\relax
\EndOfBibitem
\bibitem[Perdew \latin{et~al.}(2008)Perdew, Ruzsinszky, Csonka, Vydrov,
  Scuseria, Constantin, Zhou, and Burke]{perdew2008restoring}
Perdew,~J.~P.; Ruzsinszky,~A.; Csonka,~G.~I.; Vydrov,~O.~A.; Scuseria,~G.~E.;
  Constantin,~L.~A.; Zhou,~X.; Burke,~K. Restoring the density-gradient
  expansion for exchange in solids and surfaces. \emph{Physical review letters}
  \textbf{2008}, \emph{100}, 136406\relax
\mciteBstWouldAddEndPuncttrue
\mciteSetBstMidEndSepPunct{\mcitedefaultmidpunct}
{\mcitedefaultendpunct}{\mcitedefaultseppunct}\relax
\EndOfBibitem
\bibitem[Garrity \latin{et~al.}(2014)Garrity, Bennett, Rabe, and
  Vanderbilt]{garrity2014pseudopotentials}
Garrity,~K.~F.; Bennett,~J.~W.; Rabe,~K.~M.; Vanderbilt,~D. Pseudopotentials
  for high-throughput DFT calculations. \emph{Computational Materials Science}
  \textbf{2014}, \emph{81}, 446--452\relax
\mciteBstWouldAddEndPuncttrue
\mciteSetBstMidEndSepPunct{\mcitedefaultmidpunct}
{\mcitedefaultendpunct}{\mcitedefaultseppunct}\relax
\EndOfBibitem
\bibitem[Lee \latin{et~al.}(1995)Lee, Chang, and Cohen]{lee1995first}
Lee,~K.-H.; Chang,~K.-J.; Cohen,~M.~L. First-principles calculations of the
  Coulomb pseudopotential $\mu$*: Application to al. \emph{Physical Review B}
  \textbf{1995}, \emph{52}, 1425\relax
\mciteBstWouldAddEndPuncttrue
\mciteSetBstMidEndSepPunct{\mcitedefaultmidpunct}
{\mcitedefaultendpunct}{\mcitedefaultseppunct}\relax
\EndOfBibitem
\bibitem[Choudhary and Tavazza(2020)Choudhary, and Tavazza]{CHOUDHARY20201}
Choudhary,~K.; Tavazza,~F. In \emph{2D Nanoscale Heterostructured Materials};
  Jit,~S., Das,~S., Eds.; Micro and Nano Technologies; Elsevier, 2020; pp
  1--11\relax
\mciteBstWouldAddEndPuncttrue
\mciteSetBstMidEndSepPunct{\mcitedefaultmidpunct}
{\mcitedefaultendpunct}{\mcitedefaultseppunct}\relax
\EndOfBibitem
\bibitem[Choudhary \latin{et~al.}(2020)Choudhary, Garrity, and
  Tavazza]{Choudhary_2020}
Choudhary,~K.; Garrity,~K.~F.; Tavazza,~F. Data-driven discovery of 3D and 2D
  thermoelectric materials. \emph{Journal of Physics: Condensed Matter}
  \textbf{2020}, \emph{32}, 475501\relax
\mciteBstWouldAddEndPuncttrue
\mciteSetBstMidEndSepPunct{\mcitedefaultmidpunct}
{\mcitedefaultendpunct}{\mcitedefaultseppunct}\relax
\EndOfBibitem
\bibitem[Choudhary and Tavazza(2021)Choudhary, and
  Tavazza]{PhysRevMaterials.5.054602}
Choudhary,~K.; Tavazza,~F. Predicting anomalous quantum confinement effect in
  van der Waals materials. \emph{Phys. Rev. Mater.} \textbf{2021}, \emph{5},
  054602\relax
\mciteBstWouldAddEndPuncttrue
\mciteSetBstMidEndSepPunct{\mcitedefaultmidpunct}
{\mcitedefaultendpunct}{\mcitedefaultseppunct}\relax
\EndOfBibitem
\bibitem[Marques \latin{et~al.}(2005)Marques, L{\"u}ders, Lathiotakis, Profeta,
  Floris, Fast, Continenza, Gross, and Massidda]{marques2005ab}
Marques,~M.; L{\"u}ders,~M.; Lathiotakis,~N.; Profeta,~G.; Floris,~A.;
  Fast,~L.; Continenza,~A.; Gross,~E.; Massidda,~S. Ab initio theory of
  superconductivity. II. Application to elemental metals. \emph{Physical Review
  B} \textbf{2005}, \emph{72}, 024546\relax
\mciteBstWouldAddEndPuncttrue
\mciteSetBstMidEndSepPunct{\mcitedefaultmidpunct}
{\mcitedefaultendpunct}{\mcitedefaultseppunct}\relax
\EndOfBibitem
\bibitem[Dudarev \latin{et~al.}(1998)Dudarev, Botton, Savrasov, Humphreys, and
  Sutton]{PhysRevB.57.1505}
Dudarev,~S.~L.; Botton,~G.~A.; Savrasov,~S.~Y.; Humphreys,~C.~J.; Sutton,~A.~P.
  Electron-energy-loss spectra and the structural stability of nickel oxide: An
  LSDA+U study. \emph{Phys. Rev. B} \textbf{1998}, \emph{57}, 1505--1509\relax
\mciteBstWouldAddEndPuncttrue
\mciteSetBstMidEndSepPunct{\mcitedefaultmidpunct}
{\mcitedefaultendpunct}{\mcitedefaultseppunct}\relax
\EndOfBibitem
\bibitem[Heid \latin{et~al.}(2010)Heid, Bohnen, Sklyadneva, and
  Chulkov]{heid2010effect}
Heid,~R.; Bohnen,~K.-P.; Sklyadneva,~I.~Y.; Chulkov,~E. Effect of spin-orbit
  coupling on the electron-phonon interaction of the superconductors Pb and Tl.
  \emph{Physical Review B} \textbf{2010}, \emph{81}, 174527\relax
\mciteBstWouldAddEndPuncttrue
\mciteSetBstMidEndSepPunct{\mcitedefaultmidpunct}
{\mcitedefaultendpunct}{\mcitedefaultseppunct}\relax
\EndOfBibitem
\bibitem[Gibson and Meservey(1989)Gibson, and Meservey]{gibson1989evidence}
Gibson,~G.; Meservey,~R. Evidence for spin fluctuations in vanadium from a
  tunneling study of Fermi-liquid effects. \emph{Physical Review B}
  \textbf{1989}, \emph{40}, 8705\relax
\mciteBstWouldAddEndPuncttrue
\mciteSetBstMidEndSepPunct{\mcitedefaultmidpunct}
{\mcitedefaultendpunct}{\mcitedefaultseppunct}\relax
\EndOfBibitem
\bibitem[Sankar \latin{et~al.}(2017)Sankar, Peramaiyan, Muthuselvam, Butler,
  Dimitri, Neupane, Rao, Lin, and Chou]{zrsis-top}
Sankar,~R.; Peramaiyan,~G.; Muthuselvam,~I.~P.; Butler,~C.~J.; Dimitri,~K.;
  Neupane,~M.; Rao,~G.~N.; Lin,~M.~T.; Chou,~F.~C. Crystal growth of Dirac
  semimetal ZrSiS with high magnetoresistance and mobility. \emph{Scientific
  Reports} \textbf{2017}, \emph{7}, 40603\relax
\mciteBstWouldAddEndPuncttrue
\mciteSetBstMidEndSepPunct{\mcitedefaultmidpunct}
{\mcitedefaultendpunct}{\mcitedefaultseppunct}\relax
\EndOfBibitem
\bibitem[Aggarwal \latin{et~al.}(2019)Aggarwal, Singh, Aslam, Singha, Pariari,
  Gayen, Kabir, Mandal, and Sheet]{Aggarwal_2019}
Aggarwal,~L.; Singh,~C.~K.; Aslam,~M.; Singha,~R.; Pariari,~A.; Gayen,~S.;
  Kabir,~M.; Mandal,~P.; Sheet,~G. Tip-induced superconductivity coexisting
  with preserved topological properties in line-nodal semimetal {ZrSiS}.
  \emph{Journal of Physics: Condensed Matter} \textbf{2019}, \emph{31},
  485707\relax
\mciteBstWouldAddEndPuncttrue
\mciteSetBstMidEndSepPunct{\mcitedefaultmidpunct}
{\mcitedefaultendpunct}{\mcitedefaultseppunct}\relax
\EndOfBibitem
\bibitem[Ge \latin{et~al.}(2015)Ge, Liu, Liu, Gao, Qian, Xue, Liu, and
  Jia]{fese}
Ge,~J.-F.; Liu,~Z.-L.; Liu,~C.; Gao,~C.-L.; Qian,~D.; Xue,~Q.-K.; Liu,~Y.;
  Jia,~J.-F. Superconductivity above 100 K in single-layer FeSe films on doped
  SrTiO3. \emph{Nature Materials} \textbf{2015}, \emph{14}, 285--289\relax
\mciteBstWouldAddEndPuncttrue
\mciteSetBstMidEndSepPunct{\mcitedefaultmidpunct}
{\mcitedefaultendpunct}{\mcitedefaultseppunct}\relax
\EndOfBibitem
\bibitem[Zheng \latin{et~al.}(2018)Zheng, Gao, and Yan]{Zheng_2018}
Zheng,~D.-D.; Gao,~M.; Yan,~X.-W. Electron{\textendash}phonon coupling in
  heavily electron-doped bulk {FeSe}: a first-principles investigation.
  \emph{Applied Physics Express} \textbf{2018}, \emph{12}, 013003\relax
\mciteBstWouldAddEndPuncttrue
\mciteSetBstMidEndSepPunct{\mcitedefaultmidpunct}
{\mcitedefaultendpunct}{\mcitedefaultseppunct}\relax
\EndOfBibitem
\bibitem[Coh \latin{et~al.}(2015)Coh, Cohen, and Louie]{Coh_2015}
Coh,~S.; Cohen,~M.~L.; Louie,~S.~G. Large electron{\textendash}phonon
  interactions from {FeSe} phonons in a monolayer. \emph{New Journal of
  Physics} \textbf{2015}, \emph{17}, 073027\relax
\mciteBstWouldAddEndPuncttrue
\mciteSetBstMidEndSepPunct{\mcitedefaultmidpunct}
{\mcitedefaultendpunct}{\mcitedefaultseppunct}\relax
\EndOfBibitem
\end{mcitethebibliography}

\end{document}


\maketitle

The critical temperature of superconducting materials was determined using direct current (DC) magnetometry.  Samples of layered superconductors were purchased from Alfa Aesar (FeSe, 2H-NbSe$_2$) and 2D Semiconductors (2H-NbS$_2$ and ZrSiS).  Prior to analysis, samples were crushed and ground with a mortar and pestle to a fine powder to remove any orientation-based effects.  Magnetic susceptibility measurements were carried out using a Quantum Design Physical Property Measurement System (PPMS) equipped with a 12 T superconducting magnet.  The PPMS vibrating sample magnetometer (VSM) module was employed and the chamber was evacuated to a pressure of 10 Torr (1.33 kPa) or less.  Samples were first zero field-cooled from room temperature to 2 K whereupon a magnetic field of +0.001 T was applied.  The temperature was then raised at a rate of 1 K/min and magnetic susceptibility was measured as a function of temperature to 50 K.  A field-cooled measurement was also collected where the temperature was lowered from 50 K to 2 K at 1 K/min.  This procedure was then repeated for applied magnetic fields of +0.01 T and +0.1 T.

\begin{table}[]
\caption{q-point convergence of $\lambda$, $\omega_{\textrm{log}}$, and $T_c$ for selected promising 2D superconductors.}
\begin{tabular}{l|l|l|l|l}
\hline
Structure & q-point & $\lambda$ & $\omega_{\textrm{log}}$ (K) & $T_c$ (K)      \\
\hline
\hline
2H-TaS$_2$    & 2x2x1   & 3.3                         & 36.3                          & 6.8                          \\
          & 6x6x1   & 1.0                         & 91.3                          & 6.5                          \\
          & 7x7x1   & 1.6                         & 49.6                          & 6.1                          \\
          \hline
2H-NbSe$_2$   & 2x2x1   & 2.5                         & 46.6                          & 7.6                          \\
          & 3x3x1   & 1.4                         & 41.0                          & 4.5                          \\
          & 4x4x1   & 0.9                         & 115.8                         & 6.4                          \\
          & 6x6x1   & 0.8                         & 131.1                        & 6.7                          \\
          \hline
W$_2$N$_3$      & 2x2x1   & 0.7                         & 369.3                         & 10.6                         \\
          & 3x3x1   & 1.8                         & 102.0                         & 13.4                         \\
          & 4x4x1   & 1.4                         & 171.7                         & 18.7                         \\
          \hline
Mg$_2$B$_4$N$_2$   & 4x4x1   & 1.7 & 172.4 & 21.8 \\
          & 6x6x1   & 1.1                         & 261.2                         & 19.9 \\
          \hline
TiClO     & 4x9x1   & 2.7                         & 56.0                          & 9.7  \\
          & 6x9x1   & 2.8                         & 63.8                          & 11.1 \\
          \hline
\end{tabular}
\end{table}


\begin{figure}
\caption{Effect of different k-point and q-point selection on electron-phonon coupling (EPC) parameters for 2D 2H-NbSe$_2$ with respect to broadening: a) $\lambda$, b) $\omega_{\textrm{log}}$, and c) $T_c$. }
\begin{center}
\includegraphics[width=16cm]{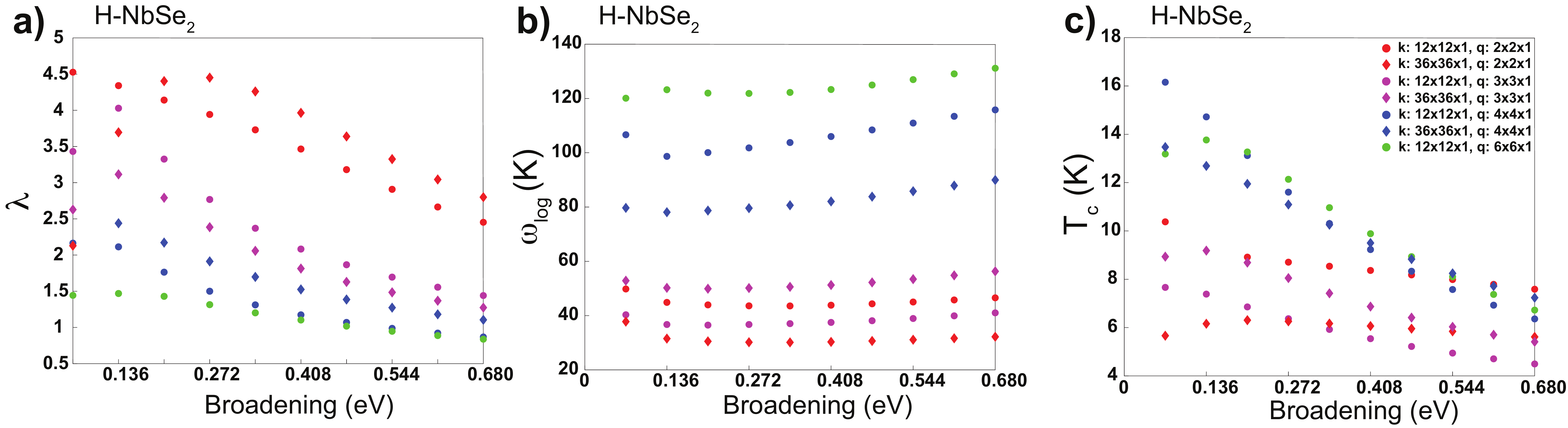}
\label{broad_q}
\end{center}
\end{figure}

\begin{figure}
\caption{The effect of broadening value for a selected group of representative superconductors: a) $\lambda$, b) $\omega_{\textrm{log}}$, c) $T_c$ and d) the change in $T_c$ with respect to $\mu^*$. }
\begin{center}
\includegraphics[width=16cm]{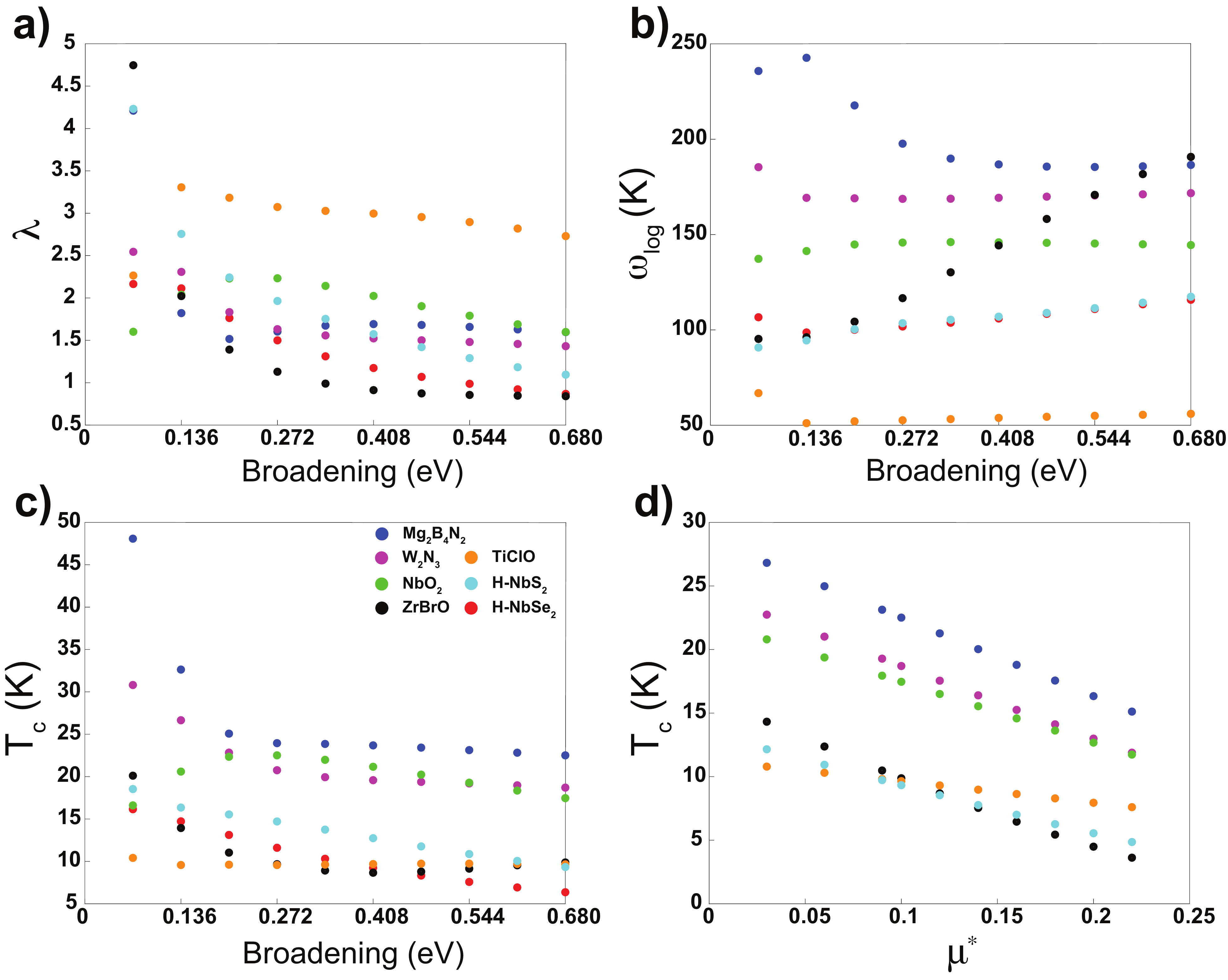}
\label{mu_broad}
\end{center}
\end{figure}

\begin{table}[]
\caption{\label{tctable} Tabulated results for the 2D superconductors with a $T_c$ above 5 K. Values for JARVIS ID, k-point grid, q-point grid, $\lambda$, $\omega_{\textrm{log}}$, $T_c$ and formation energy per atom are given. }
\begin{tabular}{l|l|l|l|l|l|l|l}
\hline
Structure & JARVIS ID          & k-point & q-point & $\lambda$ & $\omega_{\textrm{log}}$ & $T_c$ & E$_{\textrm{form}}$ \\
& & & & & (K) & (K) & (eV/atom) \\
\hline
\hline
Mg$_2$B$_4$N$_2$            & JVASP-153112 & 4x4x1            & 24x24x1          & 1.7                                 & 172.4                                 & 21.8                                                    & -0.24                              \\
W$_2$N$_3$               & JVASP-153122 & 4x4x1            & 12x12x1          & 1.4                                 & 171.7                                 & 18.7                                                    & 0.05                               \\
NbO$_2$               & JVASP-31356  & 5x5x1            & 25x25x1          & 1.6                                 & 144.5                                 & 17.5                                                    & -2.25                              \\
PtN$_2$               & JVASP-6634   & 2x2x1            & 14x14x1          & 0.8                                 & 228.1                                 & 11.9                                                    & 1.59                               \\
MgB$_2$               & JVASP-153113 & 4x4x1            & 24x24x1          & 0.6                                 & 472.2                                 & 11.8                                                    & 0.44                               \\
MoN                & JVASP-13586  & 2x2x1            & 26x26x1          & 1.1                                 & 133.7                                 & 11.2                                                    & 1.90                               \\
TiClO              & JVASP-75097  & 4x9x1            & 12x9x1           & 2.7                                 & 56.0                                  & 9.6                                                     & -2.42                              \\
B$_2$N                & JVASP-79276  & 4x4x1            & 24x24x1          & 0.7                                 & 281.2                                 & 9.6                                                     & 0.09                               \\
ZrBrO              & JVASP-28185  & 2x2x1            & 22x22x1          & 0.9                                 & 173.3                                 & 9.5                                                     & -2.39                              \\
2H-NbS$_2$             & JVASP-646    & 7x7x1            & 14x14x1          & 1.1                                 & 117.4                                 & 9.3                                                     & -1.03                              \\
NaSn$_2$S$_4$            & JVASP-6949   & 4x4x1            & 20x20x1          & 0.9                                 & 143.4                                 & 9.2                                                     & -0.60                              \\
Mg$_2$B$_4$C$_2$           & JVASP-153110 & 4x4x1            & 24x24x1          & 0.5                                 & 655.5                                 & 9.0                                                     & -0.28                              \\
Nb$_2$CuS$_4$            & JVASP-75063  & 2x2x1            & 14x14x1          & 1.5                                 & 73.1                                  & 8.6                                                     & -0.52                              \\
AuN$_2$               & JVASP-75054  & 6x4x1            & 18x32x1          & 1.4                                 & 80.6                                  & 8.4                                                     & 0.76                               \\
Nb$_2$CoS$_4$            & JVASP-27853  & 2x2x1            & 24x24x1          & 0.9                                 & 143.2                                 & 8.3                                                     & -0.59                              \\
1T-NbS$_2$             & JVASP-5947   & 4x4x1            & 24x24x1          & 1.4                                 & 79.1                                  & 8.3                                                     & -1.00                              \\
$\chi$-Borophene              & JVASP-153104 & 4x4x1            & 24x24x1          & 0.5                                 & 482.1                                 & 8.2                                                     & 0.48                               \\
NbC                & JVASP-153115 & 4x4x1            & 24x24x1          & 0.8                                 & 197.9                                 & 8.1                                                     & 0.34                               \\
ZrSiS              & JVASP-153121 & 2x2x1            & 24x24x1          & 0.9                                 & 128.2                                 & 7.9                                                     & -0.77                              \\
CoAs$_2$              & JVASP-6637   & 4x4x1            & 16x16x1          & 1.3                                 & 80.0                                  & 7.8                                                     & 0.25                               \\
$\alpha$-Mg$_2$B$_4$N$_2$            & JVASP-153111 & 4x4x1            & 24x24x1          & 1.0                                 & 104.2                                 & 7.4                                                     & 0.06                               \\
B$_2$O                & JVASP-153100 & 4x4x1            & 24x24x1          & 4.8                                 & 33.3                                  & 7.0                                                     & -0.73                              \\
2H-TaS$_2$            & JVASP-6070   & 7x7x1            & 14x14x1          & 1.0                                 & 91.3                                  & 6.5                                                     & -1.06                              \\
2H-NbSe$_2$           & JVASP-655    & 4x4x1            & 12x12x1          & 0.9                                 & 115.8                                 & 6.4                                                     & -0.76                              \\
BaSn$_4$O$_8$            & JVASP-77697  & 3x3x1            & 9x9x1            & 1.3                                 & 64.0                                  & 6.3                                                     & -1.51                              \\
1T-NbSe$_2$            & JVASP-5899   & 6x6x1            & 18x18x1          & 3.3                                 & 33.6                                  & 6.3                                                     & -0.74                              \\
LaBi$_2$O$_4$            & JVASP-28176  & 2x2x1            & 14x14x1          & 1.4                                 & 58.5                                  & 6.2                                                     & -1.92                              \\
BrCY               & JVASP-60515  & 2x2x1            & 24x24x1          & 1.0                                 & 86.8                                  & 6.0                                                     & -1.18                              \\
2H-NbTe$_2$            & JVASP-153106 & 4x4x1            & 24x24x1          & 1.1                                 & 71.8                                  & 5.8                                                     & -0.33                              \\
TiSe               & JVASP-6010   & 4x4x1            & 12x12x1          & 1.4                                 & 51.8                                  & 5.4                                                     & -0.74                              \\
TiS$_2$               & JVASP-774    & 4x4x1            & 12x12x1          & 0.7                                 & 185.4                                 & 5.4                                                     & -1.33                              \\
ZrS                & JVASP-786    & 9x9x1            & 27x27x1          & 0.8                                 & 123.3                                 & 5.3                                                     & -1.39                              \\
AuSe$_2$              & JVASP-6601   & 4x4x1            & 15x15x1          & 2.7                                 & 29.9                                  & 5.1                                                     & 0.29                               \\
VSe                & JVASP-77610  & 7x7x1            & 14x14x1          & 0.8                                 & 114.7                                 & 5.1                                                     & -0.47                             
\end{tabular}
\end{table}

\begin{figure}
\caption{Phonon density of states (DOS) for selected structures with positive formation energy: a) W$_2$N$_3$ b) MgB$_2$ c) $\chi$-Borophene d)$\alpha$-Mg$_2$B$_4$N$_2$. }
\begin{center}
\includegraphics[width=16cm]{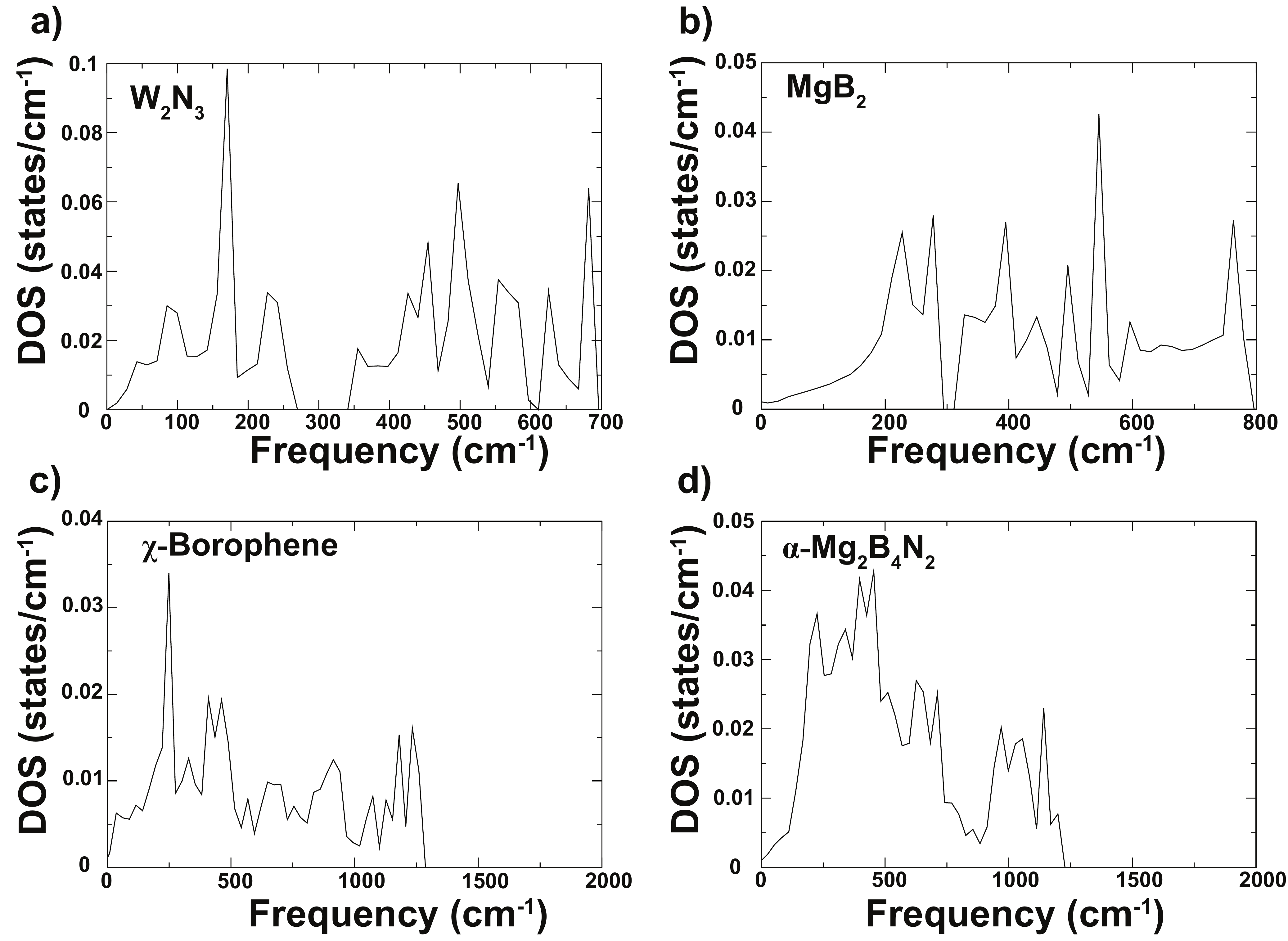}
\label{pdos}
\end{center}
\end{figure}

\begin{table}[]
\caption{\label{newstruc} The 24 newly added structures (based on literature search) to the JARVIS database with their corresponding JARVIS ID.  }
\begin{tabular}{l|l}
\hline
Structure                 & JARVIS ID    \\
\hline
\hline
Al$_4$CO & JVASP-153098 \\
$\beta$-Borophene         & JVASP-153101 \\
$\chi$-Borophene          & JVASP-153104 \\
B$_2$N                    & JVASP-153099 \\
B$_2$O                    & JVASP-153100 \\
CaC$_6$                   & JVASP-153102 \\
CaRu$_2$N$_2$             & JVASP-153103 \\
LiC$_6$                   & JVASP-153109 \\
Mg$_2$B$_4$C$_2$          & JVASP-153110 \\
Mg$_2$B$_4$N$_2$          & JVASP-153112 \\
$\alpha$-Mg$_2$B$_4$N$_2$ & JVASP-153111 \\
MgB$_2$                   & JVASP-153113 \\
MgBH                      & JVASP-153114 \\
Mo$_2$C                   & JVASP-157996            \\
NbC                       & JVASP-153115 \\
2H-NbTe$_2$               & JVASP-153106 \\
1T-NbTe$_2$               & JVASP-153118 \\
PdBi$_2$                  & JVASP-153116 \\
ScC                       & JVASP-153117 \\
2H-TaTe$_2$               & JVASP-153107 \\
1T-TaTe$_2$               & JVASP-153119 \\
W$_2$B$_3$                & JVASP-153120 \\
W$_2$N$_3$                & JVASP-153122 \\
ZrSiS                     & JVASP-153121
\end{tabular}
\end{table}
